\documentclass[reprint,superscriptaddress,aps,prb,floatfix,showkeys]{revtex4-2}
\usepackage{graphicx}
\usepackage{bm}
\usepackage{bbold}
\usepackage{amsmath}
\usepackage{amssymb}
\usepackage[version=3]{mhchem} 
\renewcommand{\vec}[1]{\bm{#1}}

\usepackage[resetlabels,labeled]{multibib}
\newcites{A}{Supplementary Material References}

\begin{document}

\author{Adriana Val\'{e}rio}
\affiliation{Institute of Physics, University of S{\~{a}}o Paulo, S{\~{a}}o Paulo, SP, Brazil}

\author{Fabiane J. Trindade}
\affiliation{Laboratory of Materials for Energy, Engineering, Modelling and Applied Social Sciences Center, Federal University of ABC, Santo Andr{\'{e}}, SP, Brazil}

\author{Rafaela F. S. Penacchio}
\affiliation{Institute of Physics, University of S{\~{a}}o Paulo, S{\~{a}}o Paulo, SP, Brazil}

\author{Bria C. Ramos}
\affiliation{Laboratory of Materials for Energy, Engineering, Modelling and Applied Social Sciences Center, Federal University of ABC, Santo Andr{\'{e}}, SP, Brazil}

\author{S{\'{e}}rgio Damasceno}
\affiliation{Laboratory of Materials for Energy, Engineering, Modelling and Applied Social Sciences Center, Federal University of ABC, Santo Andr{\'{e}}, SP, Brazil}

\author{MaurÃ­cio B. Estradiote}
\affiliation{Institute of Physics, University of S{\~{a}}o Paulo, S{\~{a}}o Paulo, SP, Brazil}

\author{Cristiane B. Rodella}
\affiliation{Brazilian Synchrotron Light Laboratory (LNLS/CNPEM), Campinas, SP, Brazil}

\author{Andr{\'{e}} S. Ferlauto}
\affiliation{Laboratory of Materials for Energy, Engineering, Modelling and Applied Social Sciences Center, Federal University of ABC, Santo Andr{\'{e}}, SP, Brazil}

\author{Stefan W. Kycia}
\affiliation{Department of Physics, University of Guelph, Guelph ON, Canada}

\author{S{\'{e}}rgio L. Morelh{\~{a}}o} 
\affiliation{Institute of Physics, University of S{\~{a}}o Paulo, S{\~{a}}o Paulo, SP, Brazil}

\title{Implications of size dispersion on X-ray small-angle scattering and diffraction of crystalline nanoparticles: ceria nanocubes as a case study}


\begin{abstract}
Small-angle X-ray scattering (SAXS) and X-ray diffraction (XRD) techniques are widely used as analytical tools in the optimization and control of nanomaterial synthesis processes. In crystalline nanoparticle systems with size distribution, the discrepant size values determined by using SAXS and XRD still lacks a well-established description in quantitative terms. To address fundamental questions, the isolated effect of size distribution is investigated by SAXS and XRD simulation in polydisperse systems of virtual nanoparticles. It quantitatively answered a few questions, among which the most accessible and reliable size values and what they stand for regarding the size distribution parameters. When a finite size distribution is introduced, the two techniques produce differing results even in perfectly crystalline nanoparticles. Once understood, the deviation in resulting size values can, in principle, resolve two parameters size distributions of crystalline nanoparticles. To demonstrate data analysis procedures in light of this understanding, XRD and SAXS experiments were carried out on a series of powder samples of cubic ceria nanoparticles. Besides changes in the size distribution related to the synthesis parameters, proper comparison of XRD and SAXS results revealed particle-particle interaction effects underneath the SAXS intensity curves. It paves the way for accurate and reliable methodologies to assess size, size dispersion, and degree of crystallinity in synthesized nanoparticles. 
\end{abstract}

\maketitle

\section{Introduction}

Metal and metal oxide crystalline nanoparticles (NPs) have been investigated for application in many research areas due to the unique physical and chemical properties related to their high surface to volume ratio \cite{vg08,jr21,yc21}. Accessing size distribution in NP systems is fundamental to understand structure-dependent functionalities, as well as morphology and other structural variables \cite{fz04,hb05,af05,ak08,ss14,cs16,gc18,ac20,ik22}. In the case of ceria (\ce{CeO2}) NPs, the reversible redox cycle (Ce$^{4+}$/Ce$^{3+}$ ratio) is associated with the formation of oxygen vacancies in the crystalline lattice having a high capacity to store and release oxygen with excellent stability \cite{cy21,yx21,cs18,mz21}. These properties afford \ce{CeO2} to be one of the most promising catalysts in many different reactions \cite{ls21,ys16,at17}. The design of catalysts by tailoring size and morphology has a significant impact on their performance. In general, hydrothermal synthesis is widely used to obtain \ce{CeO2} NPs of different morphology just by controlling critical parameters such as pH, pressure, and temperature \cite{ls21,ys16,hm05,cd20,mc20,ts19}. Despite several previous studies, the role of size dispersivity on catalytic performance are still to be analyzed.

X-ray tools such as small-angle X-ray scattering (SAXS) and X-ray diffraction (XRD) are widely used to obtain size information of NP systems \cite{tz02, ds11, ml19}. These are called bulk techniques because X-rays interact with the whole illuminated sample volume. Consequently, average values are obtained by SAXS and XRD analysis. Both techniques measure the angle-dependent distribution of the scattered radiation by atomic electrons present in the sample. The difference between them is that SAXS regime operates in scattering angles below a few degrees where it is possible to probe average size and shape of NPs regardless of crystalline perfection of the atomic structure. XRD operates at wide angles where scattering is measurable through diffraction in crystalline materials. In other words, SAXS is susceptible to the whole volume of the NPs, while XRD only accesses the crystalline volume the NPs. Combining SAXS and XRD analysis allows, in principle, optimization of synthesizing processes aimed at controlling size, size dispersion, and crystallinity of the NPs. 

Nanocrystal size information is available in the line profile of the diffraction peaks present in the XRD patterns. The so-called Scherrer equation (SE) is the bridge connecting peak width to size values in monodisperse systems \cite{ps12,ps18,ap39,uh11}. In polydisperse systems, it has been widely assumed that the width of the diffraction peaks is dictated by the volume-weighted size distribution \cite{hn00,id05,jm11,pt13}, an assumption limited to samples with narrow distributions \cite{av19}. However, there is a very long discussion that can be traced back to the early decades of the 20th century about the actual role of size distribution in the line profile of the diffraction peaks \cite{fj38,eb50,lb78,jl82,jl00,al22}. As particle dispersivity in a sample is more a consequence of its preparation history than a physical property, most traditional methods of line profile analysis are primarily concerned with effects arising from lattice imperfections and instrumentation \cite{rc92,ck98,rs99,ac00}. Available methodologies for accessing size distribution information consist of a secondary analysis of peak profiles or the use of a more elaborated approach to account for size distribution parameters when modeling diffraction patterns \cite{ps01,ps02,ac05,ml19,ps20}. In practice, attesting the accuracy and reliability of such methodologies is complicated, essentially because retrieving size distribution parameters from diffraction patterns is ambiguous due to countless size distributions capable of providing similar diffraction peaks of identical widths \cite{av19}.    

Validation of methodologies by comparing size distribution results from different techniques such as XRD and electron microscopy is also a difficult task \cite{ml04}. XRD is a bulk technique that probes the lattice coherent length,\cite{sm19} that is the crystalline domain size, on large ensemble of NPs. Electron microscopy probes shape and size distributions over a small ensemble \cite{dw09}. Both size distributions are certainly compatible in systems of fully crystalline NPs with narrow size dispersion where a small ensemble is representative of the volume sampled by XRD \cite{ps16}. In comparison with another bulk technique, such as SAXS, discrepancies in size results can arise as a consequence of NPs' crystalline domains smaller than their own sizes, as for instance in typical core/shell NPs with crystalline core and amorphous shell \cite{ri18}. In addition to this fact, attention has also to be drawn to the fact that in the SAXS regime the size distribution is weighted by the NPs' volume square \cite{ag39,ag55,ag94,ds11,sm16,jd22}. The difference in how both techniques weight the size distribution must be taken into account when comparing their size results and can be exploited to access extra information about size dispersivity and crystallinity of NPs \cite{sm22}. 

In this work, implications of size distribution in the SAXS and XRD techniques are demonstrated in polydisperse systems of virtual NPs. In such well controlled systems, free of instrumental and lattice-distortion effects, SAXS and XRD measurements can be performed to explore what quantitative information about size distribution is most accessible and reliable. Initially, total X-ray scattering patterns encompassing both SAXS and XRD are simulated via pair distance distribution function (PDDF) \cite{og77,lh12} for monodisperse sizes ranging from 1\,nm to 90\,nm. Line profile fitting of the simulated data is used to determine the accuracy by which diffraction peak width can be measured in situations of significant peak overlap, revealing relatively small uncertainties in SE constants for SAXS or XRD regimes in NPs above a few nanometers. In polydisperse systems, the obtained SE constants are used to confirm that SAXS and XRD size results correspond to the median values of the sixth and fourth moment integrals of the size distribution, respectively. SAXS and XRD measurements were carried out in real samples: a series of powder samples of ceria nanocubes. The size discrepant results from these measurements are evaluated in light of how each method weights the size distribution. Sources of additional size discrepancies and possible strategies for comparing SAXS and XRD results are discussed. Cubic morphology of most NPs and presence of size dispersion were verified by electron microscopy. Changes in the size and size dispersivity related to synthesis parameters were noticed, suggesting reliable analytical procedures to guide further optimization of chemical routes aimed at controlling these properties in crystalline ceria NPs.

\section{Theoretical Basis}

Roles of NP size and shape on SAXS and XRD processes are accurately described by using the well-known Debye scattering equation,\cite{ps16} e.g. Eq.~(\ref{eq:PofQ}), although at the high cost of computational time for NPs with sizes above a few tens of nanometers. However, as accurate quantification is needed to properly illustrate the impact of size dispersivity on both processes, the Debye scattering equation is used here as follow.

\subsection{Monodisperse systems}

A system of $N$ identical NPs randomly oriented in space scatters X-rays according to $I(Q) = N I_{\rm Th} P(Q)$ where $I_{\rm Th}$ is the Thomson scattered intensity by a single electron accounting for polarization effects and
\begin{equation}\label{eq:PofQ}
    P(Q) = \sum_a\sum_b f_a(Q) f_b^*(Q) \frac{\sin(Qr_{ab})}{Qr_{ab}}
\end{equation}
is the NP scattering power \cite{pd15,ag94,lg10,sm16,ps16}. $Q=(4\pi/\lambda)\sin\theta$ is the modulus of the scattering vector for a $2\theta$ angle of scattering. $f_{a,b}(Q)$ is the atomic scattering factor of atom $a$ or $b$ with resonant amplitudes for X-ray of wavelength $\lambda$. $r_{ab} = |\vec r_a - \vec r_b|$ are the distances between any pair of atoms at the $\vec r_a$ and $\vec r_b$ frozen-in-time positions; indices $a$ and $b$ run over all the atoms in the NP.

Temporal coherence of X-ray beams is of the order of 1\,fs. At this time scale, atomic thermal vibrations are very slow. Actual diffraction patterns acquired over longer periods of time correspond to the sum of instantaneous intensity of statistically equivalent structures with displaced atoms within their amplitudes of vibration. Powder samples of identical NPs stand for ensembles of such statistically equivalent structures. Random uncorrelated displacements are commonly accounted for in the Debye-Waller factors \cite{ps16}, while collective vibrations---phonons---are responsible for thermal diffuse scattering \cite{an11}. For the purposes of this work, random displacements $d\vec r$ around the average position $ \langle {\vec r}_a \rangle $ were considered for each atom as in ${\vec r}_a = \langle {\vec r}_a \rangle + d\vec r$. The scattering power of a frozen NP with random displacements is practically the same of the average scattering power computed over the ensemble of statistically equivalent structures; see a demonstration in Supporting Information S1.

At small scattering angles where resolution is not enough to resolve the shortest atomic distances inside the NP, the scattering power

\begin{eqnarray}\label{eq:intpofu}
    P(Q) & = & 4\pi \int c(u) u^2 \frac{\sin(Qu)}{Qu} du = \nonumber \\ & = &\int p(u)\frac{\sin(Qu)}{Qu} du.
\end{eqnarray}
has been written in terms of the electron-electron pair correlation function $c(u)=\rho_s(r)*\rho_s(-r)$ \cite{ag94}, equivalent to the Patterson function for crystals\cite{ap35}; $\rho_s(r)$ is a spherically symmetric electron density function representing the NP in the system of identical NPs randomly oriented in space \cite{sm16}. In the case of wide-angle scattering, by comparing Eqs.~(\ref{eq:PofQ}) and (\ref{eq:intpofu}), it is easy to see that
\begin{equation}\label{eq:pofu}
   p(u) = \sum_a\sum_b f_a(Q) f_b^*(Q) \delta(u - r_{ab})
\end{equation}
can be calculated as atomic scattering factor weighted PDDFs; $\delta()$ is the Dirac delta function. Each PDDF is a histogram between chemical species for which the real value of $f_a(Q) f_{b}^\ast(Q)$, that is  $\Re\left[f_a(Q) f_{b}^\ast(Q)\right]$, is a common factor \cite{lg10}. To optimize computational time, Eq.~(\ref{eq:pofu}) can be rewritten as

\begin{eqnarray}\label{eq:pofu2}
p(u) & = & \sum_\alpha |f_\alpha(Q)|^2H_{\alpha}(u) + \nonumber \\ 
    & + & 2\sum_\alpha\sum_\beta \Re\left[f_\alpha(Q) f_{\beta}^\ast(Q)\right] H_{\alpha\beta}(u)
\end{eqnarray}
where $\alpha$ and $\beta$ run over the number of different chemical species in the NP.
\begin{equation}
    H_{\alpha} (u) = N_\alpha \delta(u) + 2\sum_{a = 1}^{N_\alpha} \sum_{b>a}^{N_\alpha} \delta (u - r_{ab})\,,
\label{Eq:HaDeU}
\end{equation}
is the histogram of distances between atoms $a$ and $b$ of a same chemical specie,  and
\begin{equation}
    H_{\alpha\beta} (u) = \sum_{a = 1}^{N_\alpha} \sum_{b=1}^{N_\beta} \delta (u - r_{ab})
\label{Eq:HabDeU}
\end{equation}
is the histogram of atomic distances where atoms $a$ and $b$, belong to the sets of $N_\alpha$ and $N_\beta$ atoms of the different chemical species $\alpha$ and $\beta$, respectively.

For NPs of a given size and shape, that is a monodisperse system, simulated total scattering patterns embracing both SAXS and XRD processes are obtained by the scattering power $P(Q)$ in Eq.~(\ref{eq:intpofu}), after computing the above PDDFs in Eqs.~(\ref{Eq:HaDeU}) and (\ref{Eq:HabDeU}) \cite{lg10,ac15}.

\subsection{Polydisperse systems}

In systems of non-identical monocrystalline NPs diffracting independently from each other, that is dilute systems where no spatial correlation exist between the NPs \cite{tz02} and secondary diffraction processes are negligible \cite{sm91}, the diffracted X-ray intensities can be described by
\begin{equation}\label{eq:IofQ}
    I(Q) = I_{\rm Th}\int P_k(Q) n(k) dk
\end{equation}
where $n(k)dk$ stands for the number of NPs sharing a common property with value in the range $k$ and $k+dk$. The scattering power $P_k(Q)$ of the NPs in this range follows from Eq.~(\ref{eq:intpofu}), as long as they form a complete set of randomly oriented NPs. Here, the variable $k$ refers to the size of the NPs, implying the assumption that they all have the same shape. For instance, $k$ stands for the edge length $L$ in systems of cubic NPs, or the diameter $D$ for spherical NPs.

According to the kinematical theory of X-ray diffraction, all atoms of a given system of particles are subjected to the same incident wave, and the scattered waves thus produced suffer no further interaction with the system \cite{bw90,ag94}. Under this kinematical approach, the integrated intensity $\mathcal{I}_k = \int P_k(Q) dQ$ around each one of the diffraction peaks observed in $P_k(Q>0)$ is proportional to the NP volume, that is $\mathcal{I}_k \propto k^3$. On the other hand, while the peak area increases with the NP volume, the peak width $\mathcal{W}_k$ gets narrower inversely with the NP size $k$, that is $\mathcal{W}_k \propto k^{-1}$. Consequently, the peak maximum height at position $Q_0$ goes as $P_k(Q_0)\propto k^4$. Besides the diffraction peaks, there is also the scattering peak around the direct beam at $Q=0$, that is the SAXS peak. Eq.~(\ref{eq:PofQ}) provides the squared number of scattering electrons in this case \cite{ag55,ag94}, that is proportional to the squared volume. Hence, the scattering peak maximum height goes as $P_k(0)\propto k^6$. 

In samples with a particle size distribution (PSD) function $n(k)$, the resulting peak  $\mathcal{W}$, that is, the peak full width at half maximum (FWHM), follows from Eq.~(\ref{eq:IofQ}) as
\begin{eqnarray}\label{eq:Ihalf}
 I(Q_0\pm \mathcal{W}/2) & = & \frac{1}{2} I_{\rm Th} \int P_k(Q_0)n(k)dk = \nonumber \\
   & = &  I_{\rm Th} \int_0^{\widetilde{K}_m} P_k(Q_0)n(k)dk\,,
\end{eqnarray}
clearly establishing that $\mathcal{W}$ is related to the median value $\widetilde{K}_m$ of the PSD function weighted by $P_k(Q_0)$. Hence, $\widetilde{K}_m$ is the median value of the $k^m$-weighted PSD, that is
\begin{equation}\label{eq:tildeK}
    \int_0^{\widetilde{K}_m} k^m n(k)dk = \frac{1}{2} \int_0^\infty k^m n(k)dk
\end{equation}
where $m=4$ for all diffraction peaks, and $m=6$ for the small-angle scattering peak. In $Q$ space, the widely known SE (Scherrer equation) is simply $k\,\mathcal{W}_k = \Upsilon$ in the case of monodisperse systems with X-ray patterns as given by $P_k(Q)$, Eq.~(\ref{eq:intpofu}). The SE constant $\Upsilon$ is determined by the NP shape, may differ from one diffraction peak to another in the case of non-spherical NPs, and it is always slightly different for the small-angle scattering peak even for spherical NPs. In the case of polydisperse systems with a general PSD, the same SE constants apply.\cite{sm22} A measure of the peak width $\mathcal{W}$ leads to the median value
\begin{equation}\label{eq:SE}
    \widetilde{K}_m = \frac{\Upsilon}{\mathcal{W}} = \frac{\Upsilon}{2\pi}\frac{\lambda}{\cos\theta_0\,\mathcal{W}_{2\theta}}
\end{equation}
when estimating the NP size in the sample via the SE. The SE in Eq.~(\ref{eq:SE}) is also written in its most used form where $\mathcal{W}_{2\theta}$ is the peak width as a function of the scattering angle $2\theta$ given that $\theta_0 = \arcsin\left( \frac{\lambda}{4\pi}Q_0\right)$ is the Bragg angle for diffraction peaks and it is zero for the small-angle scattering peak. The $\Upsilon/2\pi$ ratio is commonly called shape factor of the SE, whose value is close to 1; hence, $\Upsilon \simeq 2\pi$. In this work, the $\Upsilon$ values for the small-angle scattering and diffraction peaks will be extracted from the X-ray patterns of virtual NPs, as they will be used for size measurements in simulated patterns of polydisperse systems. Nonetheless, similar values can also be obtained from available SAXS and XRD literature \cite{tz02,jl78,ps01}.

\section{Materials and Methods}

\subsection{Virtual nanoparticles}

Vibrational disorder was introduced in the virtual NPs by adding $[ dx,\,dy,\,dz] = \delta r [ \zeta_1,\,\zeta_2,\,\zeta_3]$ to the mean atomic positions $\langle \vec{r}_a\rangle = [X_a,\,Y_a,\,Z_a]$ as computed for perfectly periodic lattices. The random numbers $\zeta_n$ are in the range $[0,1]$, and the atomic disorder parameter $\delta r$ produces a Gaussian broadening of width $\delta r$ in the histograms of pair distances, corresponding to a nearly isotropic root-mean-square (RMS) displacement $\langle dr \rangle_{\rm rms} = \delta r/(4\sqrt{\ln{2}\,}) \simeq 0.3\delta r$, as shown in Supporting Information S1. For each virtual NP, the histograms in Eqs.~(\ref{Eq:HaDeU}) and (\ref{Eq:HabDeU}) have been calculated with a bin of $\Delta u = 0.002$\,\AA, as in $H(u) = \sum_a \sum_b \int_u^{u + \Delta u} \delta(u' - r_{ab}) du'$. Computer codes were written in C$++$ for parallel processing mode with 36 cores. Execution environment was a virtual machine running Debian-amd64 8.11 operating system (56 cores and 16GB Mem), hosted by a HPE ProLiant DL360 Gen10 Server with Xen kernel, 64 cores of Intel(R) Xeon(R) Gold 5218 CPU @ 2.30GHz / 256 GB Mem. For a NP with $N_{\rm at}$ atoms of a same chemical specie, the observed time to compute $H_\alpha(u)$, Eq.~(\ref{Eq:HaDeU}), is about $50\,N_{\rm at}^{2.151}$\,ps. For instance, a NP with $N_{at} = 19\times10^6$, having $N_{at}(N_{at}-1)=361\times10^{12}$ pair distances to be computed, takes a processing time of 63 hours. The atomic scattering factors  $f(Q)=f_0(Q)+f^{\prime}(\lambda)+if^{\prime\prime}(\lambda)$ in Eq.~(\ref{eq:pofu2}) have been calculated by routines \texttt{asfQ.m} and \texttt{fpfpp.m}, both available at \citeauthor{mc16}(\citeyear{mc16}). The scattering power of the NPs were then obtained from the integral in Eq.~(\ref{eq:intpofu}), which  becomes a discrete sum of the $p(u)$ values on each histogram bin weighted by the corresponding $\frac{\sin(Qu)}{Qu}$ factor, that is $P(Q) = \sum_n p(u_n) \frac{\sin(Qu_n)}{Qu_n}$ where $u_{n+1} = u_n + \Delta u$.

\subsection{Ceria nanocubes}

\ce{CeO2} nanocubes were synthesized via a hydrothermal process by adaptation of a well-established protocol \cite{hm05}. Two different concentrations 6\,M, sample labelled B5, and 12\,M, samples labelled B11 and C1, of sodium hydroxide (NaOH, P.A-A.C.S. 100\%, Synth) was dissolved in 35\,mL of deionized (DI) water. 0.1\,M of cerium nitrate (Ce(NO$_3$)$_3$.6H$_2$O,  $\geq 99.9$\%, Sigma-Aldrich) precursor was dissolved in 10\,mL of DI water and added to NaOH solution with stirring. For sample C1, 0.025\,M of urea (\ce{CH4N2O}, P.A-A.C.S. 100\%, Synth) was also added as a structure-directing agent. After stirring for 15\,min, the slurry solution was transferred into a 100\,mL stainless steel vessel autoclave and heated at 180$^\circ$C in an electric oven for 24\,h, and allowed to cool at room temperature. The final product was collected by centrifugation, followed by washing in DI water and  ethanol, five times in each. Finally, the precipitate was dried in an electric oven at 80$^\circ$C for 12\,h and ground into a powder.

\subsection{Electron microscopy analysis}
Scanning electron microscopy (SEM) images were obtained with JEOL microscopes operating at 5\,kV. Samples were prepared by drop-casting of nanostructure aqueous suspensions over double stick carbon tape, drying under ambient conditions, and sputtering of 3\,nm platinum coating to improve the signal-noise ratio.

\subsection{XRD analysis}
X-ray diffraction (XRD) analysis of \ce{CeO2} powder samples were performed in a D8 Discover Bruker with LYNXEYE XE-T detector. Cu$K_\alpha$ radiation (1.5418\,\AA), Bragg Brentano $\theta$-$\theta$ goniometry, and step-scanning mode: step size of 0.015$^\circ$ and counting time of 2s per step. Sample spinning velocity was 15 rpm. A NIST corundum 676a standard\cite{jc11} at room temperature has been used to determine the instrumental broadening  $\mathcal{W}_{\rm inst} = (0.355 Q + 1.81)\times10^{-3}$ \,\AA$^{-1}$ of this diffractometer as a function of $Q$, Supporting Information S2. Diffraction peak widths, without instrumental broadening, were determined as $\mathcal{W}  = (\mathcal{W}_{\rm exp}^2 - \mathcal{W}_{\rm inst}^2)^{1/2}$
where $\mathcal{W}_{\rm exp}$ stands for the actual FWHM extracted by line profile fitting with pseudo-Voigt functions, as described in Supporting Information S2. The codes were written in MatLab, similar to the one used in a previous work \cite{ac20} where accuracy in determining the FWHM was a critical issue.

\subsubsection{SAXS analysis}
SAXS data were acquired in a Xeuss 2.0 system (Xenocs, France), sourced by a microfocus GeniX$^{\rm 3D}$ X-ray generator, Cr$K_\alpha$ radiation (2.2923\,\AA), followed by a FOX$^{\rm 3D}$ collimating mirror, and ultra-high resolution mode where beam cross-section is set to $0.4\times0.4\,{\rm mm}^2$ and the Pilatus 300K (Dectris, Switzerland) area detector placed at 6.46\,m from the sample. X-ray longitudinal coherence length of 310\,nm is limited by the spectral line width of the characteristic radiation, while the transverse coherence length is determined by X-ray collimation and estimated to approach 500\,nm by the equipment manufacturer. Total acquisition time per sample was 1\,hour. The intensities scattered on the detector area have been converted into curves as a function of $Q$ by using the homemade software available at the Center of Multi-User SAXS Equipment of the Institute of Physics, University of S{\~{a}}o Paulo, Brazil, where the measurements were carried out.  Background scattering from kapton films, used to hold the powder samples in the vacuum chamber, have been subtracted from the intensity curves. Minimum $Q$ value considered for the \ce{CeO2} samples was $1.83\times10^{-3}\,\textrm{\AA}^{-1}$, for which no beam stopper shadow correction is required.

\section{Computational Results}

\begin{figure*}
    \centering
    \includegraphics[width=.75\textwidth]{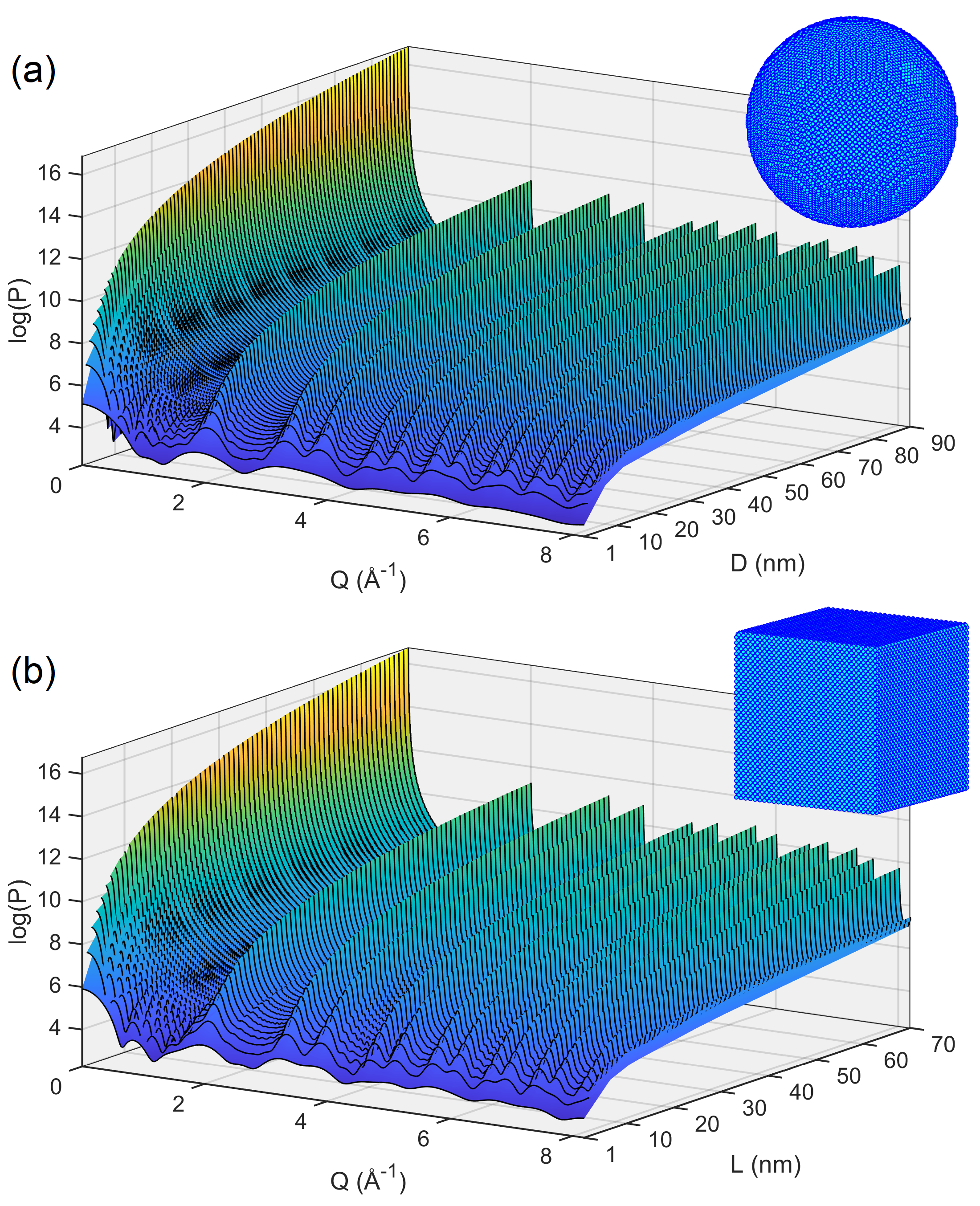}
    \caption{Scattering power $P(Q)$, Eq.~(\ref{eq:intpofu}), obtained by PDDF calculation in virtual NPs of silicon. (a) Spherical NPs of diameter $D$, ranging from 1 to 90\,nm, and (b) cubic NPs of edge $L$, ranging from 1 to 70\,nm. A small vibrational disorder $\delta r$ of 2\% regarding the shortest pair distance (Si-Si bond length, $235.16$\,pm), that is a RMS displacement $\langle dr \rangle_{\rm rms} = 1.41$\,pm, have been applied to all atomic sites in the NPs. Examples of spherical ($D = 20$\,nm) and cubic ($L = 16$\,nm) virtual NPs with about $2\!\times\!10^5$ atoms each are shown in the insets.}
    \label{plot3DxrdsiliconnanoDL}
\end{figure*}

Fig.~\ref{plot3DxrdsiliconnanoDL} displays the $P(Q)$ curves of spherical and cubic silicon NPs as standing for X-ray total patterns of monodisperse systems. Single element NPs were considered to optimize computer time. Even so, the PDDFs of the largest NPs, containing between 17 and 19 million atoms, took more than 48 hours to compute as previously detailed (\S\, Materials and Methods). The two sets of simulated patterns with sizes covering the widest possible range serve three purposes: \textit{i)} to verify values and uncertainties of the SE constants for crystallites of nanoscale sizes; \textit{ii)} to simulate size dispersion by composing weighted patterns based on these monodisperse patterns; and \textit{iii)} to evaluate at least two distinct NP shapes of which in one the diffraction peak widths are susceptible to the crystallographic orientation of the NP facets. Weighted patterns and verified SE constants are needed to accurately quantify SAXS and XRD size results in polydisperse systems of virtual NPs.

\begin{figure}
    \centering
    \includegraphics[width=.35\textwidth]{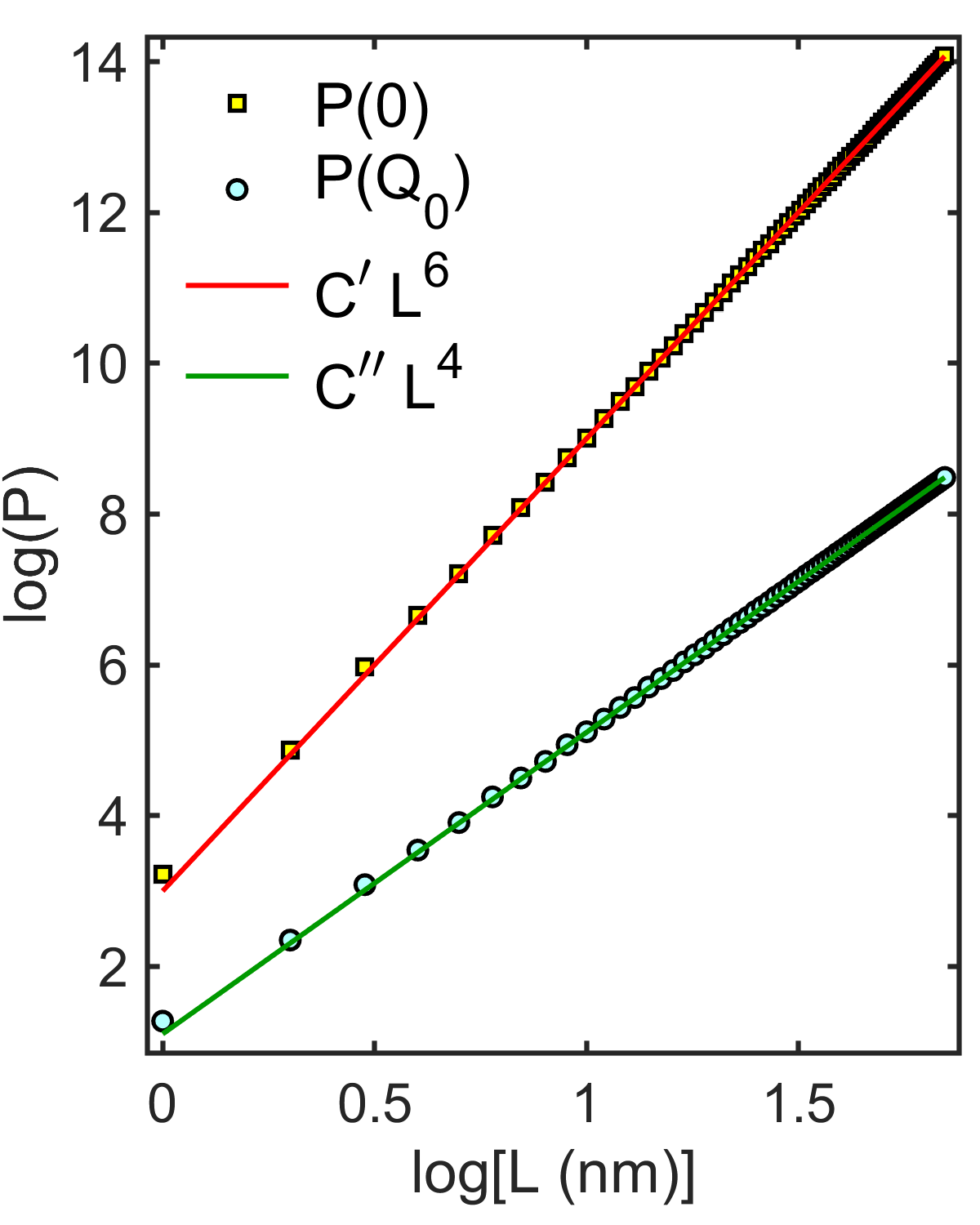}
    \caption{SAXS (squares) and XRD (circles) peak maximum heights as a function of NP cube edge length $L$, as obtained from the $P(Q)$ curves in Fig.~\ref{plot3DxrdsiliconnanoDL}(b) for $Q=0$ and $Q_0 = 3.2723\,\textrm{\AA}^{-1}$, respectively. Expected behaviour (solid lines) of the maxima according to the X-ray kinematical theory, that is $P(0) = C^{\prime}L^6$ and $P(Q_0) = C^{\prime\prime} L^4$, are also shown; the proportionality constants adjusted values are $C^\prime = 1003$ and $C^{\prime\prime} = 12.77$.}
    \label{plotk6k4}
\end{figure}

Fig.~\ref{plotk6k4} compares the maximum heights of the scattering peak $P(0)$ and of one diffraction peak $P(Q_0>0)$ as a function of the cube edge $L$, showing that they are proportional to $L^6$ and $L^4$, respectively. Similar plots are obtained for spherical NPs as a function of the diameter. According to Eq.~(\ref{eq:tildeK}), knowing the exact behavior of peak height versus size is important because it establishes the proper moment integral of the particle size distribution function.

\begin{figure}
    \centering
    \includegraphics[width=0.35\textwidth]{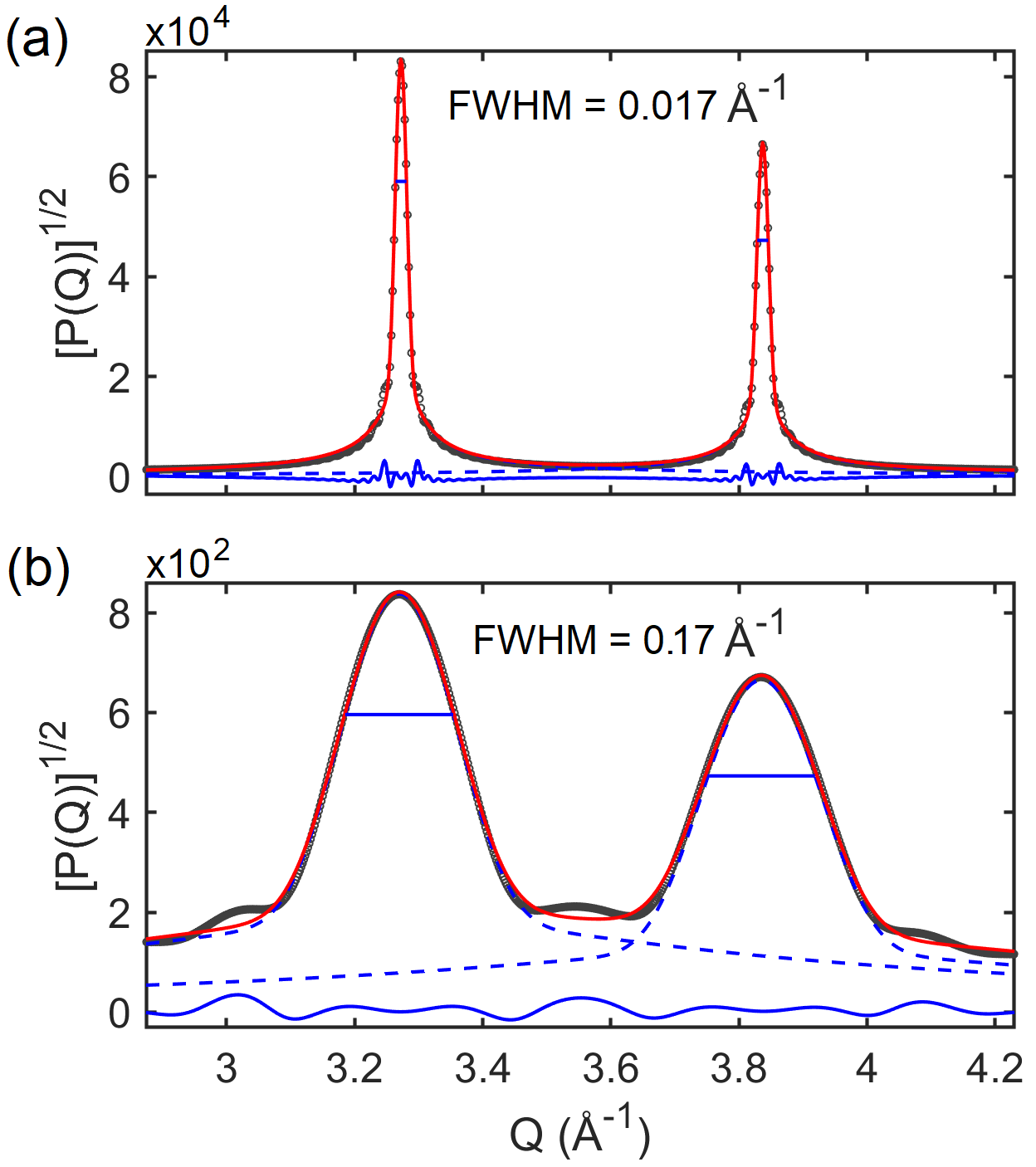}
    \caption{Nearby XRD peaks of monodisperse NPs with two distinct diameters: (a) $D=40$\,nm and (b) $D=4$\,nm. Simulated peaks (black open circles), line profile fitting (red lines) with two pseudo-Voigt functions (blue dashed lines), and residue analysis (blue line). A factor 10 in diameter ratio leads to factors of $10^{-1}$ in width and $10^4$ in height of the peaks.}
    \label{fig:lpfit400x40}
\end{figure}

\begin{figure*}
    \centering
    \includegraphics[width=.75\textwidth]{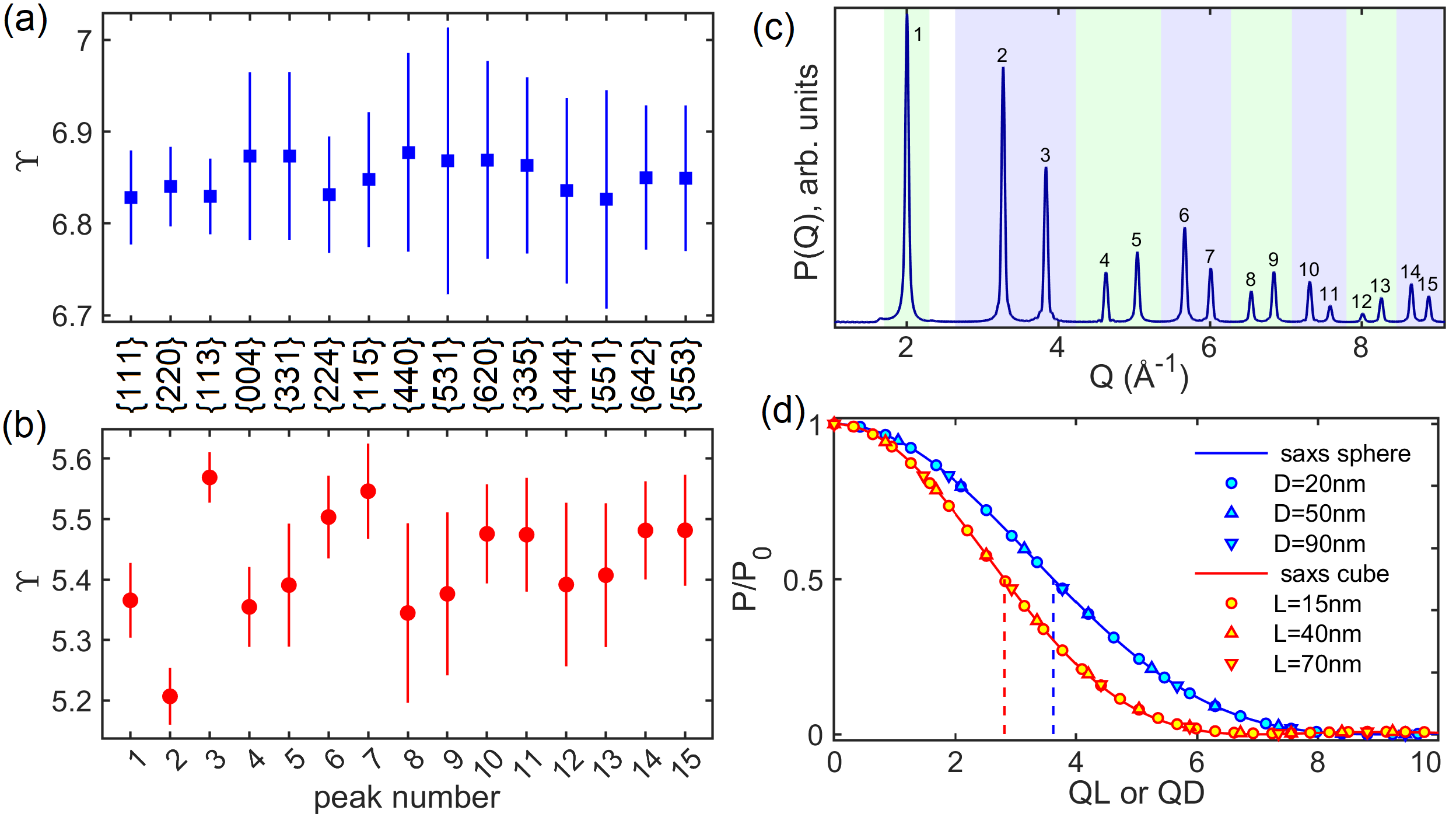}
    \caption{(a,b) SE constant $\Upsilon$ obtained from the monodisperse patterns in Fig.~\ref{plot3DxrdsiliconnanoDL} for (a) spherical NPs and (b) cubic NPs with (001) facets. Vertical bars are the standard deviation values, computed over the ensemble of NPs with sizes above 3\,nm. (c) XRD pattern, that is the $P(Q>1\,\textrm{\AA}^{-1})$ curve, for a NP of edge $L=10$\,nm, showing numerical identifiers and intervals of fitting (shaded areas of different colors) of the peaks used to determine the constant $\Upsilon$. The Bragg reflections family contributing to each peak is indicated in between the (a) and (b) panels. (d) SAXS curves of a few NPs (symbols) plotted against $QD$ or $QL$, and compared with the line profile functions (solid lines) for SAXS intensity curves, Eq.~(\ref{eq:saxslpfunc}). The half height for sphere and cubes are, respectively, at $QD= 3.630$ and $QL = 2.822$ (dashed lines). }
    \label{figuresummary}
\end{figure*}

For NPs with size smaller than 4\,nm, diffraction peak overlapping is too severe, as can be seen in Fig.~\ref{plot3DxrdsiliconnanoDL} for $D$ or $L < 4$\,nm. Above this size, line profile fitting with pseudo-Voigt functions has been applied to measure the diffraction peak width. Examples of peak fitting are given in Fig.~\ref{fig:lpfit400x40} for spherical NPs of diameter $D$, further confirming that diffraction peak intensities scale up with $D^4$ while peak width goes with $D^{-1}$. A summary of mean $\Upsilon$ values and their standard deviations obtained by fitting the first fifteen diffraction peaks of $P(Q)$ curves is given in Fig.~\ref{figuresummary}(a,b). Peak numbers and intervals of fitting are indicated in Fig.~\ref{figuresummary}(c). For spherical NPs, all values fall within the standard deviation bars of each other, while for cubic NPs the mean values can differ by more than one standard deviation. By using $\Upsilon_{D} = 6.84\pm0.05$ for spherical NPs and $\Upsilon_{L} = 5.4\pm0.2$ for cubic NPs the minimum uncertainties in diffraction-based size determination through the SE are about 1\% and 4\%, respectively. These are the minimum uncertainties as there are also the uncertainties from peak width measurements. The uncertainty $\pm0.2$ in $\Upsilon_{L}$  takes into account all variations in peak width due to crystallographic orientation of NP facets, allowing size estimation without knowing the orientation of the NP facets in the samples.

The SE constant $\Upsilon_0$ for the SAXS peak is slightly different of the SE constants $\Upsilon$ for the XRD peaks. Its value is more accurate (uncertainty below 0.1\%) than diffraction ones, as it can be determined without overlapping effects of nearby peaks. $\Upsilon_{0D}=7.260$ for spheres and $\Upsilon_{0L}=5.643$ for cubes as shown in Fig.~\ref{figuresummary}(d), where the computed $P(Q)$ curves of the NPs at low $Q$ are compared with line profile functions of half width $\Upsilon_{0D}/2D$ and $\Upsilon_{0L}/2L$. Both line profile functions have the general form

\begin{equation}\label{eq:saxslpfunc}
    \Phi(x) = [\sin(x)-x \cos(x)]^2/x^6
\end{equation}
where $x = \frac{1}{2}QD$ for spheres of diameter $D$, \cite{ag55} and $x = \frac{1}{2}\left(\Upsilon_{0D}/\Upsilon_{0L}\right) QL$ for cubes of edge $L$. Spheres and cubes have the same volume when $D/L=1.241$, a close value to the obtained ratio $\Upsilon_{0D}/\Upsilon_{0L}=1.286$. All values of the SE constants for the SAXS curve and XRD peaks are presented in Table~\ref{tab:SEconstants}.

\begin{table}\label{tab:SEconstants}
\begin{center}
\caption{SE constants for SAXS ($\Upsilon_0$) and XRD ($\Upsilon$) peaks as determined from the monodisperse X-ray patterns in Fig.~\ref{plot3DxrdsiliconnanoDL}. Theoretical diffraction SE constants (Theo. Diff.) are also shown. $^\dag$Mean value and standard deviation for 52 main reflections in a cubic crystallite  \cite{jl78}. }
\begin{tabular}{crrr}
\hline\hline
{\small NP shape} & {\small SAXS} & {\small XRD} & {\small Theo. Diff.} \\
\hline
Sphere & $\Upsilon_{0D}=7.260$ & $\Upsilon_{D} = 6.84$ & $6.7525$ \\
 & $\pm0.002$ & $\pm 0.05$ & \\
 \hline
Cube & $\Upsilon_{0L}=5.643$ & $\Upsilon_{L} = 5.4$ & $5.53\,^\dag$\\
 & $\pm0.002$ & $\pm 0.2$ & $\pm0.14$ \\
\hline\hline
\end{tabular}
\end{center}
\end{table}

\begin{figure*}
    \centering
    \includegraphics[width=.75\textwidth]{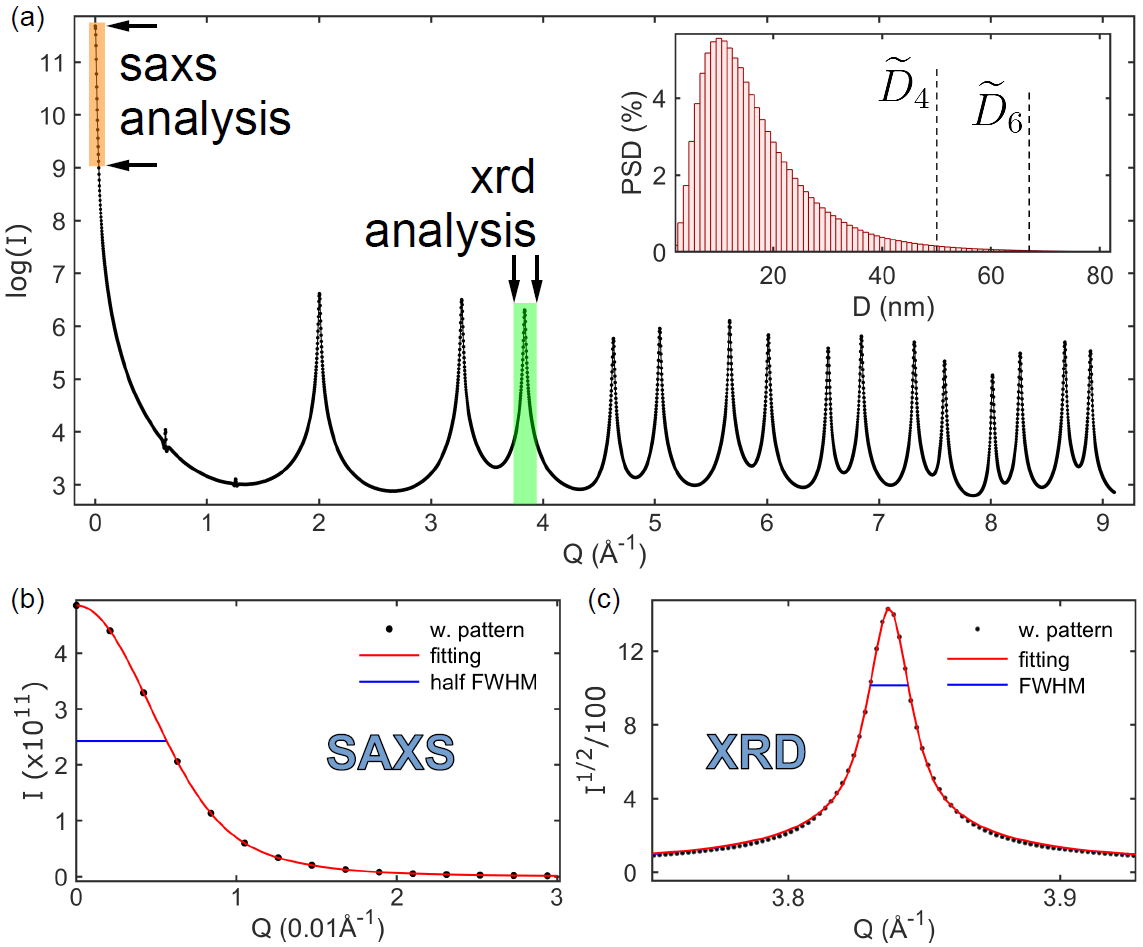}
    \caption{(a) Simulated X-ray pattern (dots with line) of polydisperse spheres given in number of scattering electrons, that is $I(Q)$ in Eq.~(\ref{eq:IofQ}) with $I_{\rm Th}=1$. (Inset) Weighting function for the monodisperse X-ray patterns in Fig.~\ref{plot3DxrdsiliconnanoDL}(a), taken as a PSD function with diameter median values $\widetilde{D}_4$ and $\widetilde{D}_6$. Intervals in $Q$ for SAXS and XRD size analysis are indicated by shaded areas and arrows. (b) SAXS peak of the weighted pattern (dots) and curve fitting (solid red line) based on analytical expression of SAXS intensity curve for spheres, Eq.~(\ref{eq:saxslpfunc}). (c) Diffraction peak of the weighted pattern (dots) and line profile fitting (solid red line) based on pseudo-Voigt function. Heights at half maximum of the scattering and diffraction peaks are indicated (horizontal blue lines). Peak width (FWHM): $\mathcal{W}_{\rm saxs} = 11.3\times10^{-3}\,\textrm{\AA}^{-1}$ and $\mathcal{W}_{\rm xrd} = 14.1\times10^{-3}\,\textrm{\AA}^{-1}$.}
    \label{figurePSDanalysis}
\end{figure*}

Total X-ray intensity patterns of polydisperse samples carry two distinct pieces of information about the size distribution. As SAXS and XRD intensity maxima have heights proportional to the NP sizes raised to different powers, 6 and 4 respectively, peak width measurements provide median values of the sixth and fourth moment integrals of the PSD, as indicated in Eq.~(\ref{eq:tildeK}). To demonstrate this direct correlation between peak widths and median values of the PSD moment integrals, the computed $P(Q)$ curves for monodisperse sizes in Fig.~\ref{plot3DxrdsiliconnanoDL} can be used to simulate the effect of polydispersivity in X-ray patterns by choosing a particular PSD function. For instance, Fig.~\ref{figurePSDanalysis}(a) presents the weighted pattern of a polydisperse system of spheres with discrete lognormal PSD function of 1\,nm size increment in the range from 1\,nm to 90\,nm. The corresponding diameter median values for the PSD moment integrals are $\widetilde{D}_{6}=66.1$\,nm and $\widetilde{D}_{4}=49.4$\,nm, inset of Fig.~\ref{figurePSDanalysis}(a). Note that both values are much larger than the 10\,nm value of the PSD mode. Details of SAXS and XRD analysis of the weighted pattern are shown in Figs.~\ref{figurePSDanalysis}(b) and \ref{figurePSDanalysis}(c), respectively. The small-angle scattering peak of half width $\mathcal{W}_{\rm saxs}/2 = \Upsilon_{0D}/2D_{\rm saxs} = 5.643\times10^{-3}\,\textrm{\AA}^{-1}$ provides the SAXS size result $D_{\rm saxs} = 64.3$\,nm. On the other hand, the diffraction peak of width $\mathcal{W}_{\rm xrd} = \Upsilon_{D}/D_{\rm xrd} = 1.410\times10^{-2}\,\textrm{\AA}^{-1}$ provides the XRD size result $D_{\rm xrd} = 48.5$\,nm.  Within the 1\,nm accuracy of this numerical demonstration, the obtained size results perfectly agree with the corresponding median values of the PSD moment integrals. In Table~\ref{tab:psdanalysis}, median values and size results of weighted patterns for narrower PSDs are also shown. Besides the good agreement between these values, Table~\ref{tab:psdanalysis} (fourth column) also reports that the fraction $x$ of Gaussian contribution in the pseudo-Voigt line profile fitting decreases as the size dispersion increases. Larger the size dispersion, more Lorentzian-like are the diffraction peaks. Therefore, independently of changes in the aspect of the line profile function due to size distribution, the SE constants for monodisperse systems are also valid for polydisperse sizes. It establishes a bridge to connect the SAXS and XRD peak widths with the median values of the weighted size distribution, following a general property of peak functions as pointed out elsewhere \cite{sm22}.

\begin{table}
\begin{center}
\caption{Comparison of median values, $\widetilde{D}_{4}$ and $\widetilde{D}_{6}$, with simulated XRD and SAXS size results, $D_{\rm xrd}$ and $D_{\rm saxs}$, as a function of the size dispersion parameter $\sigma$. Numerical values, accurate to within 0.5\,nm, are based on simulated polydisperse patterns for spheres, as the one in Fig.~\ref{figurePSDanalysis}(a). The PSDs were taken as discrete 1\,nm bin lognormal functions of standard deviation $\sigma$, 10\,nm mode, and having 90\,nm as the maximum size. Parameter $x$ stands for the fraction of Gaussian contribution in the pseudo-Voigt line profile fitting of the XRD peaks (Supporting Information S2).}\label{tab:psdanalysis}
\begin{tabular}{cccccc}
\hline\hline
$\sigma$ & $\widetilde{D}_{4}$ (nm) & $D_{\rm xrd}$ (nm) & $x$ & $\widetilde{D}_{6}$ (nm) & $D_{\rm saxs}$ (nm) \\
\hline
0.1 & 10.1 & 10.6 & 0.93 & 10.3 & 10.8 \\
0.2 & 11.7 & 12.2 & 0.84 & 12.8 & 13.4 \\
0.4 & 21.8 & 22.6 & 0.69 & 30.1 & 31.8 \\
0.6 & 49.4 & 48.6 & 0.25 & 66.1 & 64.3 \\
\hline\hline
\end{tabular}
\end{center}
\end{table}

\newpage
\section{Experimental Results}

\begin{figure*}
    \centering
    \includegraphics[width=.75\textwidth]{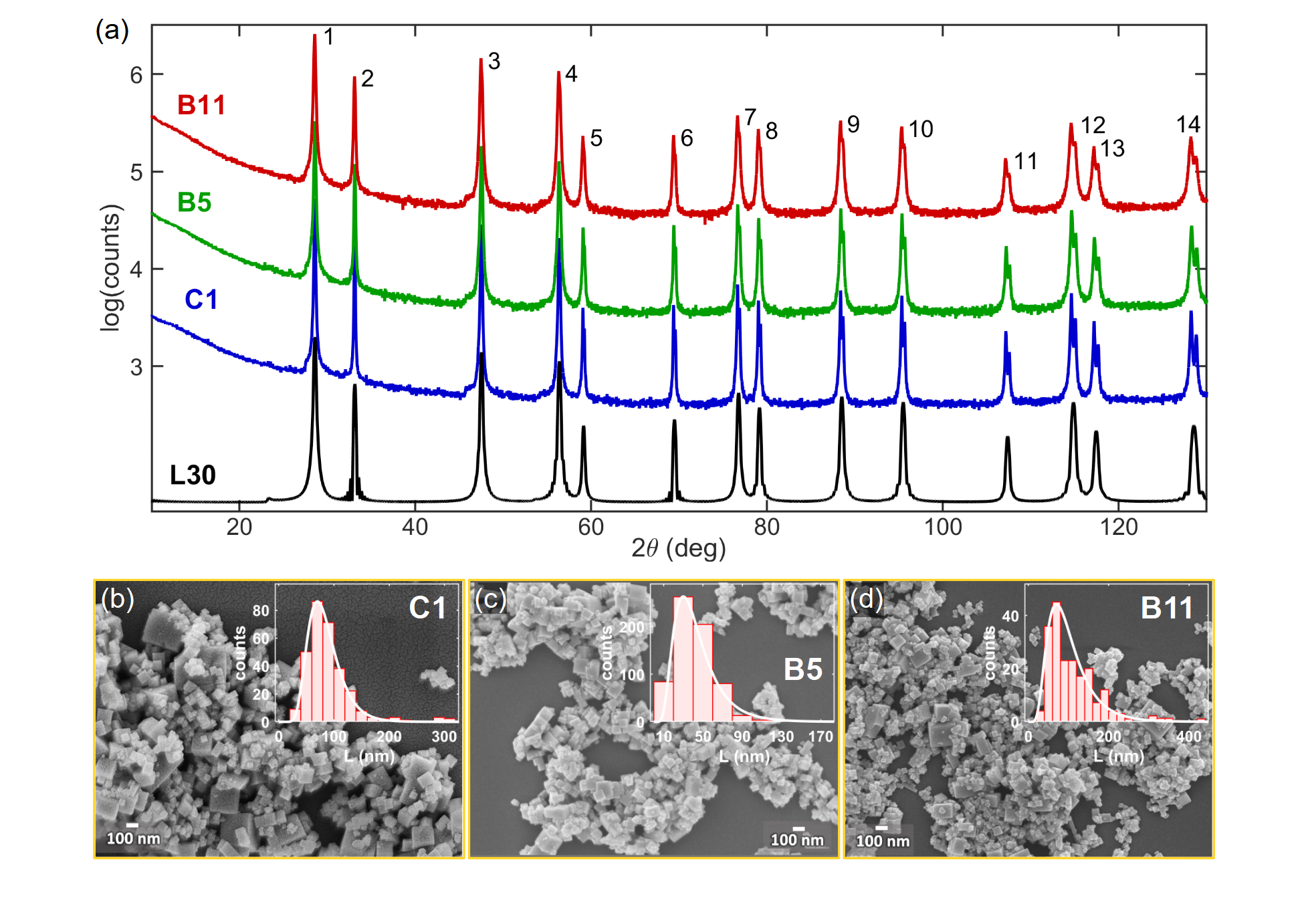}
    \caption{(a) XRD patterns of powder samples C1, B5, and B11 of CeO$_2$ NPs with Cu$K_\alpha$ radiation. A simulated pattern (L30) for monodisperse system of cubic NPs of edge $L=30$\,nm is also shown as reference, Cu$K_{\alpha 1}$ only, RMS displacement $\langle dr \rangle_{\rm rms} = 1.4$\,pm as vibrational disorder of about 2\% regarding the shortest pair distance (Ce\textendash O bond length, $234$\,pm). (b-c) SEM images and histogram (inset) of apparent size distribution for each powder sample as indicated. }
    \label{fig:plotB5B11C1}
\end{figure*}

\begin{figure*}
    \centering
    \includegraphics[width=.75\textwidth]{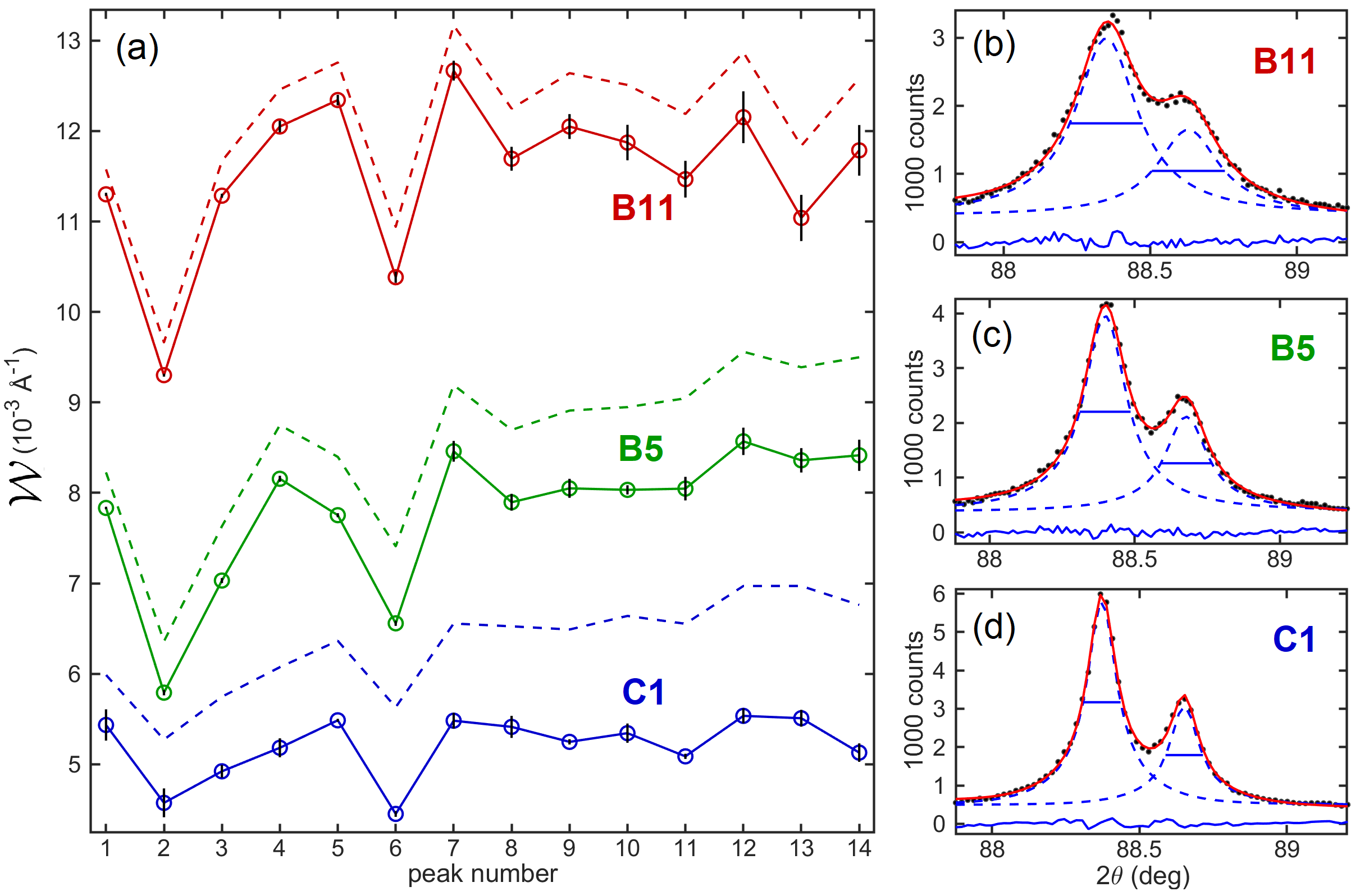}
    \caption{(a) Experimental peak width $\mathcal{W}$ (open circles connected by lines) from XRD line profile analysis in CeO$_2$ cubic NPs, samples B11, B5, and C1. Vertical error bars stand for difference in peak width from $K\alpha_1$ and $K\alpha_2$ lines. Peak widths before deconvolution of instrumental broadening are also indicated (dashed blue lines). (b-d) Examples of line profile fitting (solid red lines) of the peak number 9 (scatter black dots) with pseudo-Voigt functions (dashed blue lines); see Supporting Information S2 for the other peaks.}
    \label{fig:fwhmB5B11C1}
\end{figure*}

XRD patterns and SEM images of the powder samples of CeO$_2$ are shown in Fig.~\ref{fig:plotB5B11C1}. According to the images, all samples have cubic NPs with sizes ranging from about ten to hundreds of nanometers, Fig.~\ref{fig:plotB5B11C1}(c-d). The apparent size distributions are highly dependent on the ensemble of NPs in each image. On the other hand, XRD patterns clearly reveal differences in the median values of particle sizes from one sample to another. It is seen qualitatively by the greater or lesser overlap of the $K_{\alpha1}$ and $K_{\alpha2}$ diffraction peaks at higher angles, as in peaks 12 to 14 in Fig.~\ref{fig:plotB5B11C1}(a), see also the zoom of peak 9 in Fig.~\ref{fig:fwhmB5B11C1}(b-d), suggesting the largest diffracting crystallites are present in sample C1 and the smallest ones in sample B11. Quantitative analyzes of peak widths based on pseudo-Voigt line profile fitting with $K_\alpha$ radiation (Supporting Information S2) are shown in Fig.~\ref{fig:fwhmB5B11C1}(a). For cubic NPs, the edge $L_{\rm xrd} = \Upsilon_{L}/\mathcal{W}_{\rm xrd}$ stands for the XRD size result. The peak width is different for each reflection, but they are clearly different for each sample, as notice in Fig.~\ref{fig:fwhmB5B11C1}(a). In order not to underestimate the uncertainties, the widths $\mathcal{W}_{\rm xrd}=5.3\pm0.2\times10^{-3}\,\textrm{\AA}^{-1}$ (sample C1), $8.0\pm0.4\times10^{-3}\,\textrm{\AA}^{-1}$ (sample B5), and $11.8\pm0.5\times10^{-3}\,\textrm{\AA}^{-1}$ (sample B11) are used for size measure, taken as the average width values regarding the 14 diffraction peaks with standard deviations (not the standard deviation of the mean) as uncertainty. By using $\Upsilon_{L}=5.4\pm0.2$, the XRD size results of the samples were obtained as presented in Table~\ref{tab:results} (column 2). The NPs were considered strain free as the peak widths show no systematic variation as a function of $Q$. 

The SE constant $\Upsilon_{L}=5.4\pm0.2$, Table~\ref{tab:SEconstants}, obtained for cubic NPs of silicon is also valid for the ceria nanocubes, as well as for cubic NPs of any other compound. To make this point clear, the simulated XRD pattern for ceria NPs of edge $L=30$\,nm in Fig.~\ref{fig:plotB5B11C1}(a) has average width $\mathcal{W}_{\rm xrd}=18.0\pm0.4\times10^{-3}\,\textrm{\AA}^{-1}$ that provides $L_{\rm xrd}=30.0\pm1.3$\,nm as the XRD size result, in agreement with the size and shape of the virtual NP used to compute the  Ce\textendash Ce, Ce\textendash O, and O\textendash O pair distance histograms in Eqs.~(\ref{Eq:HaDeU}) and (\ref{Eq:HabDeU}).

\begin{figure}
    \centering
    \includegraphics[width=0.35\textwidth]{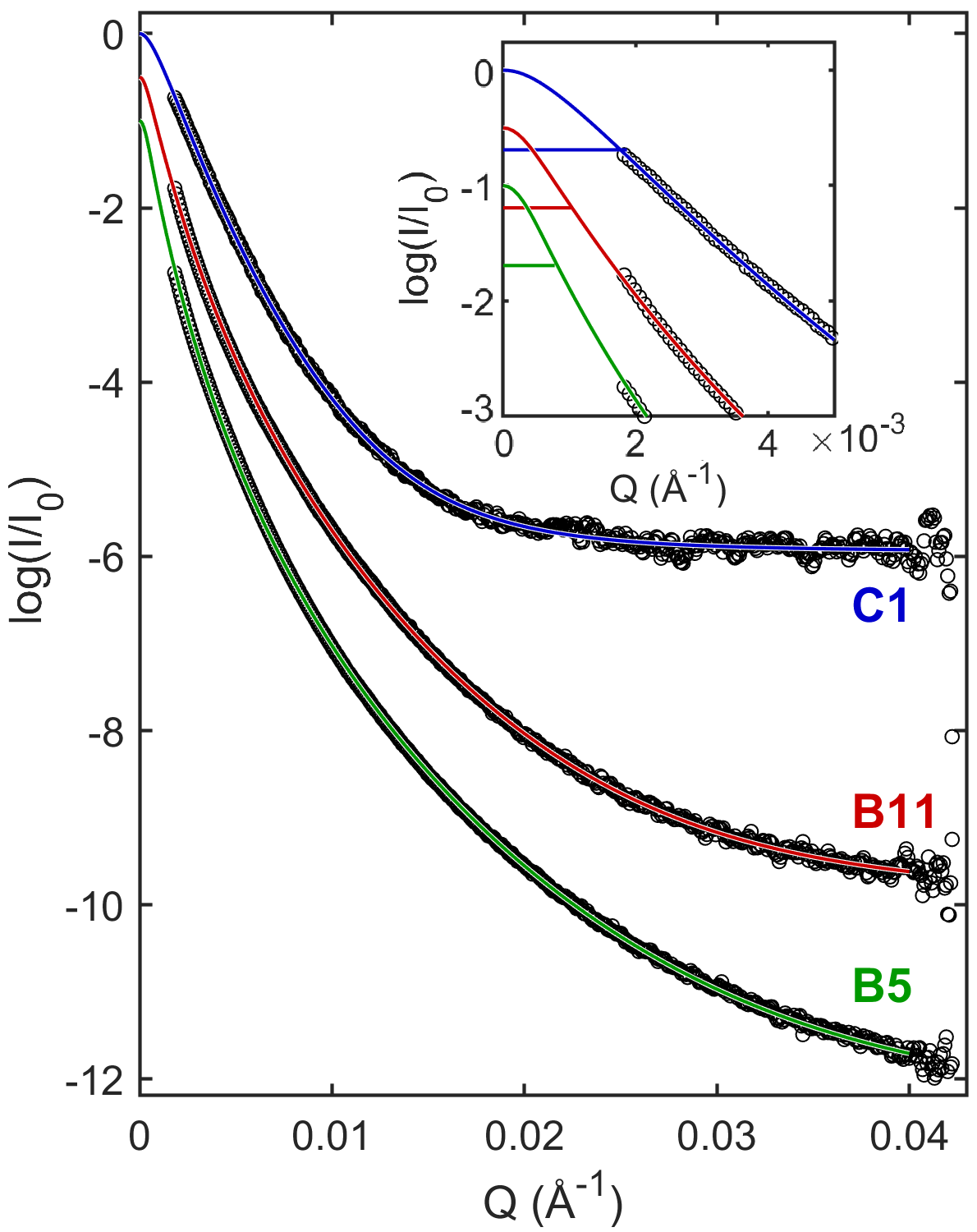}
    \caption{Experimental (open circles) and adjusted (solid lines) SAXS intensity curves from \ce{CeO2} cubic NPs, powder samples B5, B11, and C1. $I/I_0$ stands for normalized intensity. Adjusted curves were obtained through the $Y(Q)$ fitting function in Eq.~(\ref{eq:YofQ}). Ultra-small $Q$ region of the curves are highlighted as inset where the height at half-maximum of each adjusted curve is indicated (horizontal solid lines). }
    \label{fig:saxsB5B11C1}
\end{figure}

Fig.~\ref{fig:saxsB5B11C1} shows the SAXS measurements of samples C1, B11, and B5, as well as the best fit curves of the data. As the intensity at $Q=0$ is inaccessible in the used SAXS equipment, the scattering peak half widths $\mathcal{W}_{\rm saxs}/2 = \Upsilon_{0L}/2L_{\rm saxs} = 1.77\pm0.04\times10^{-3}\,\textrm{\AA}^{-1}$ (sample C1), $1.04\pm0.02\times10^{-3}\,\textrm{\AA}^{-1}$ (sample B11), and $0.82\pm0.03\times10^{-3}\,\textrm{\AA}^{-1}$ (sample B5) were estimated by extrapolating the fitted curves to $Q=0$, inset of Fig.~\ref{fig:saxsB5B11C1}. The uncertainties were estimated by repeating the fitting procedure several times after introducing statistical noise into each data points. By using $\Upsilon_{0L}=5.64$, Fig.~\ref{figuresummary}(d), the SAXS size results of the samples were obtained as presented in Table~\ref{tab:results} (column 3).

\begin{table*}
\begin{center}
\caption{Size results $L_{\rm xrd}$ and $L_{\rm saxs}$, as determined by XRD and SAXS experiments in the C1, B11, and B5 powder samples. Lognormal size distribution parameters $\sigma$ and  $L_0$ as obtained by Eq.~(\ref{eq:sigmaL0}), or by SAXS data fitting via Eq.~(\ref{eq:YofQ}). Median $\widetilde{L}_m=L_0\exp[(m+1)\sigma^2]$ as the expected XRD ($m=4$) and SAXS ($m=6$) size results from SAXS perspective. Last digits uncertainties are in parentheses.}\label{tab:results}
\begin{tabular}{cccc|cc|cccc}
\hline\hline
 & & & &\multicolumn{2}{c|}{from Eq.~(\ref{eq:sigmaL0})} & \multicolumn{4}{c}{from 
 Eq.~(\ref{eq:YofQ})}\\
sample & $L_{\rm xrd}$ & $L_{\rm saxs}$ & $\Delta L/L$ & $\sigma$ & $L_0$ & $\sigma$ & $L_0$ & $\widetilde{L}_4$ & $\widetilde{L}_6$ \\
 & (nm) & (nm) &  &  & (nm) & & (nm) & (nm) & (nm) \\
\hline
 C1 & 102(5) &  160(4) & 0.36 & 0.44(5) & 38(7) & 0.623(4) & 10.0(2) & 70(3) & 151(6) \\
 B5 & 67(4) & 345(13) & 0.81 & 0.92(4) & 1.0(3) & 0.788(6) & 4.7(2) & 105(6) & 364 (27) \\
B11 &  46(3) &  272(6) & 0.83 & 0.94(3) & 0.6(1) & 0.768(3) &  4.3(1) & 82(3) & 267(10) \\
\hline\hline
\end{tabular}
\end{center}
\end{table*}

\section{Discussions}

To interpret as adequately as possible discrepant XRD and SAXS size results, such as $L_{\rm xrd}$ and $L_{\rm saxs}$ in Table~\ref{tab:results}, it is necessary to first evaluate possible discrepancies for actual size distributions as if there is no interaction between NPs and their sizes have exactly the coherent lengths of their crystal lattices, that is, considering a dilute system of perfect crystalline NPs. This evaluation requires choosing an appropriately PSD function. In the particular case of one of the most common probability functions for size distribution, which is the lognormal function \cite{ck98,lk99,hj06}, the $m^{th}$ moment integral has analytical solution $\widetilde{L}_m = L_0\exp[(m+1)\sigma^2]$ for its median value given in terms of the most probable size (mode) $L_0$ and standard deviation $\sigma$ (Supporting Information S3). By taking the medians $\widetilde{L}_4$ and $\widetilde{L}_6$ as the XRD and SAXS size results respectively, the corresponding size distribution parameters follow from 
\begin{eqnarray}\label{eq:sigmaL0}
\sigma^2 & = & -\frac{1}{2}\ln{(L_{\rm xrd}/L_{\rm saxs})}\quad{\rm and}\quad L_0 = \nonumber \\ & = & L_{\rm xrd}\exp(-5\sigma^2) = L_{\rm saxs}\exp(-7\sigma^2)\,.
\end{eqnarray}
It shows that the discrepancy $\Delta L/L = 1-L_{\rm xrd}/L_{\rm saxs} = 1-\exp(-2\sigma^2)$ in size results arises as an exclusive consequence of the size dispersion parameter $\sigma$ under the idealized conditions above mentioned. The contribution of partial crystallization to the size discrepancy is taken into account here within the assumption that the volume fraction $\xi$ between the crystalline domain and the whole NP volume is nearly constant with respect to the NP size. Then, the size discrepancy accounting for both effects becomes 
\begin{equation}\label{eq:DL_L}
    \Delta L/L = 1-L_{\rm xrd}/L_{\rm saxs} = 1-\xi^{1/3}\,\exp(-2\sigma^2)\,.
\end{equation}
In samples with narrow size distributions and crystalline NPs, as those represented by simulated patterns with $\sigma\leq0.2$ in Table~\ref{tab:psdanalysis} (1st and 2nd rows), no detectable discrepancy is introduced for error bars larger than 5\% in the XRD and SAXS size values. In other words, compatible size results, that is $\Delta L/L<0.08$ [from Eq.~(\ref{eq:DL_L}) with $\xi=1$ and $\sigma=0.2$], only occurs for systems of crystalline NPs with narrow size distributions. For example, SAXS curves displaying well-defined fringes are typical signatures of monodisperse sizes \cite{tz02,lw18}. In such cases, measuring $L_{\rm xrd}\approx L_{\rm saxs}$ stands as a proof that the NPs are also highly crystalline, $\xi>0.78$ [from Eq.~(\ref{eq:DL_L}) with $\Delta L/L<0.08$ and $\sigma=0$].   

Eq.~(\ref{eq:sigmaL0}) imposes no upper limit to the dispersion parameter $\sigma$. However, lognormal functions with $\sigma>0.8$ represent distributions with very broad size dispersion that can range from a few nanometers to near a micrometer. By stipulating this $\sigma$ value as the upper limit of physically acceptable size dispersivity when synthesizing NPs, discrepancies where $\Delta L/L > 0.72$ can be taken as a strong evidence that there are other effects contributing to the observed discrepancy value. According to this criterion, only sample C1 shows a discrepancy $\Delta L/L = 0.36$ small enough to be totally explained in terms of size distribution. When it is the case, the parameters $\sigma$ and $L_0$ of the distribution can be directly obtained from Eq.~(\ref{eq:sigmaL0}) as presented in Table~\ref{tab:results} (columns 5 and 6). But, for samples B5 and B11, the size discrepancy $\Delta L/L \geq 0.81$ is too large, suggesting that SAXS is probing NP sizes much larger than their crystal lattices.   

From the SAXS perspective only, the size distribution parameters were already determined when fitting the SAXS curves. The analytical expression in Eq.~(\ref{eq:saxslpfunc}) adapted for cubes of edge $L$, weighted by $L^6$, and by a lognormal function $n(L)$, led to a parametric SAXS fitting function
\begin{equation}\label{eq:YofQ}
    Y(Q) = a\int L^6 \Phi(Q,L) n(L) dL + b
\end{equation}
with four adjustable parameters: normalization factor $a$, constant background intensity $b$, and the lognormal parameters $L_0$ and $\sigma$. Data fittings were driven by a genetic algorithm that minimizes the mean-square error of the log transformed data.\cite{mw99} The best fits of the SAXS data are those shown in Fig.~\ref{fig:saxsB5B11C1}, while the $L_0$ and $\sigma$ values characterizing the size distributions from the SAXS perspective are given in Table~\ref{tab:results} (columns 7 and 8). Note that the median values $\widetilde{L}_6$ for these distributions match the $L_{saxs}$ values from the half width measurements. Without XRD measurements, all available information about the samples' size dispersivity have to rely on the fitting procedures of the SAXS curves in Fig.~\ref{fig:saxsB5B11C1}. As the fitting quality is very good for the analyzed samples, no room is left to question the reliability of the SAXS results or the used methodology based on polydisperse sizes of non-interacting cubic NPs. The scenario is completely different when results from another bulk technique as XRD are introduced. For comparison purposes, the size distribution parameters $\sigma$ and $L_0$ from SAXS curve fitting are used to calculate the expected XRD size result $\widetilde{L}_4$ (Table~\ref{tab:results}, column 9). 

\begin{figure*}
    \centering
    \includegraphics[width=.75\textwidth]{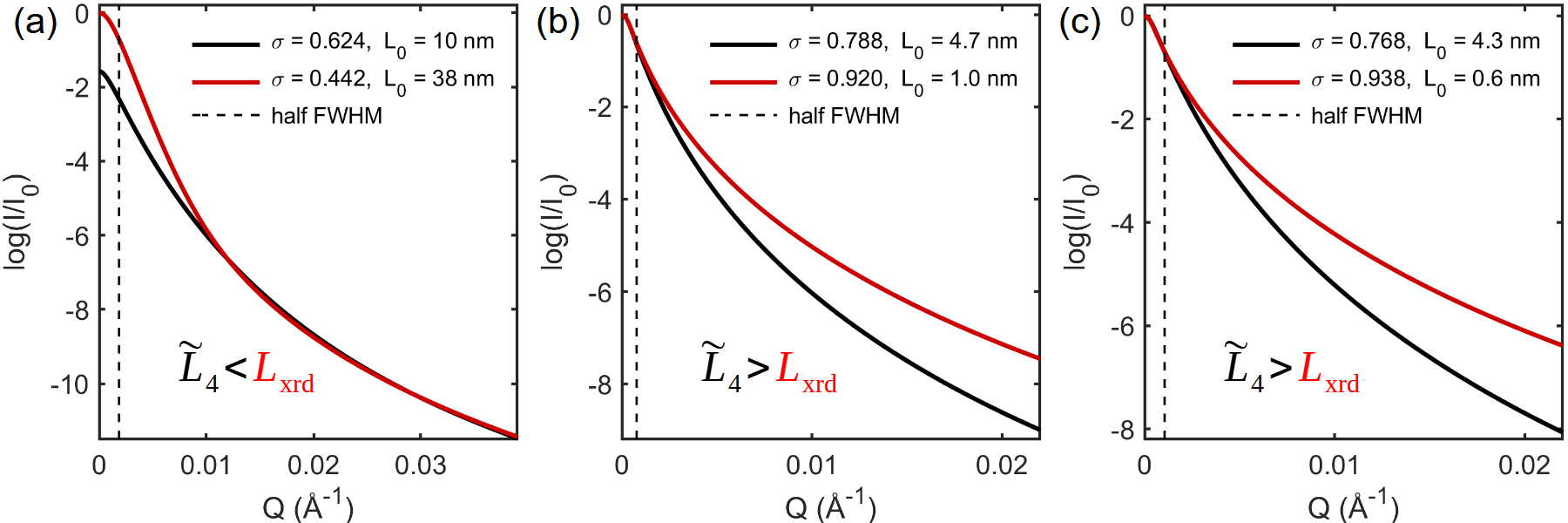}
    \caption{Simulated SAXS curves of polydisperse systems of cubic NPs with lognormal size distribution parameters $\sigma$ and $L_0$ from Table~\ref{tab:results}, as indicated. In each panel, the curves have very close widths and $\widetilde{L}_6$ values, but different values for $\widetilde{L}_4$. $L_{\rm xrd}$ in Table~\ref{tab:results} is the $\widetilde{L}_4$ value for one of the curves. (a) Curves with different intensity maxima at $Q=0$ and similar asymptotes; one curve is displaced in the vertical axis for the sake of comparison. (b,c) Curves with different asymptotes and similar intensities around $Q=0$.}
    \label{fig:simsaxsplot}
\end{figure*}

For sample C1, $\widetilde{L}_4 < L_{\rm xrd}$. It shows that the dispersion parameter $\sigma = 0.623$ optimized by curve fitting leads to the physical inconsistency of NP crystalline domain larger than the NP itself. This result is important, as it illustrates the relevance of combining XRD and SAXS size results to evaluate the reliability of methodologies based on a single bulk technique. A possible cause of this inconsistency is an intensity reduction centered in $Q=0$, see Fig.~\ref{fig:simsaxsplot}(a), characteristic of highly packed systems of NPs with similar sizes where the scattering becomes susceptible to the NP-NP exclusion distance (NPs impenetrability) \cite{tz02,sm16}. 

For samples B5 and B11, $\widetilde{L}_4 > L_{\rm xrd}$ is a physically feasible situation where SAXS probes NP sizes larger than their crystalline domains. In ideally dilute systems where the NP-NP distances are larger than the coherence length of the incident X-rays (\S\, Materials and Methods), the possible explanation goes towards NPs of partial crystallization. When it is the case, an estimate of the crystalline volume fraction $\xi = (L_{\rm xrd}/\widetilde{L}_4)^3$ is possible within the assumptions that led to Eq.~(\ref{eq:DL_L}). Then, $26\pm6$\% and $18\pm4$\% are the volume fractions of crystalline material in the NPs of samples B5 and B11, respectively. However, powder samples are in general far from dilute systems, especially when the NPs appear very clustered in the SEM images, Fig.~\ref{fig:plotB5B11C1}(c,d). It opens opportunity to also consider other effects, such as samples with broad size dispersion of crystalline NPs where a great number of small particles wrap around the larger ones, blurring particles' interfaces from SAXS perspective and producing an apparent size effect of much larger particles. Particles interface properties are known to affect the SAXS curve asymptotic behavior, as in the Porod's law \cite{sc88}. The cause of $\widetilde{L}_4 > L_{\rm xrd}$ is the difference in asymptotic behaviour, Figs.~\ref{fig:simsaxsplot}(b,c), supporting the hypothesis that SAXS is seeing NP aggregates instead of individual NPs. 

Proper interpretation of the XRD size result is a crucial point when comparing results from both techniques, and it can be highlighted here by taking the results for sample B11 as an example. The $L_{\rm xrd}$ value accessed via XRD peak width through the SE, Eq.~(\ref{eq:SE}), stands for the median value $\widetilde{L}_4$ of the actual $L^4$-weighted size distribution. The median $\widetilde{L}_3 = L_0\exp(4\sigma^2)$ of the volume-weighted size distribution can be much closer to the actual XRD size result, but $\widetilde{L}_3$ is not to be compared to $L_{\rm xrd}$ for the following reasons. Sample B11 has $\widetilde{L}_3= 46$\,nm according to the SAXS fitting results in Table~\ref{tab:results} (columns 7 and 8), a value that perfectly matches the XRD size result of this sample, $L_{\rm xrd}=46\pm3$\,nm in Table~\ref{tab:results} (column 2). It shows that, by misinterpreting the XRD size as $\widetilde{L}_3$, size dispersion alone can explain the XRD and SAXS size discrepancy, leaving no room to question the SAXS fitting procedure, NP-NP interaction effects, and NP crystallinity. In other words, the size discrepancy leads to narrower size dispersion, and the sample appears more dilute and crystalline when $L_{\rm xrd}$ is mistaken by $\widetilde{L}_3$.   

By comparing chemical routes from the cerium nitrate precursor towards the synthesized samples B5 ({\small NaOH 6M}), B11 ({\small NaOH 12M}), and C1 ({\small NaOH 12M + \ce{CH4N2O} 0.025M}), it is evident that sodium hydroxide acts as nucleation agent while urea is able to slow down nucleation and favor crystal growth, which eventually narrows size distribution around larger sizes as seen for sample C1. The small size discrepancy ($\Delta L/L = 0.36$) reported for this sample suggests highly crystalline NPs. In synthesizing samples B5 and B11, the higher the concentration of sodium hydroxide the smaller the XRD size result, indicating its action mostly as a nucleation agent. Assertive discussions about the actual size dispersivity and crystallinity of CeO$_2$ NPs were hidden by their spontaneous self-aggregation observed in the series of samples analyzed here, making unfeasible the preparation of dilute samples for proper SAXS measurements. Reliable analytical procedures for further controlling size, size dispersivity, and crystallinity is possible by comparing SAXS and XRD results in light of how these bulk techniques weight the size distribution, although the feasibility of preparing dilute samples has also to be a concern when synthesizing the NPs. Besides dilute systems, it is also relevant to emphasize that ultra-SAXS instrumental configurations capable of actually measuring the scattering peak around the direct beam, as those configurations based on analyzer crystals,\cite{dc97,ep03,aa06,lr07,sm10} improve the reliability on SAXS size results as the peak maximum height and therefore its width are available measures irrespective of curve fitting approaches. To assertively evaluate size dispersivity and crystallinity of NPs in advanced X-ray instruments capable of performing both techniques simultaneously \cite{ji18,as21,os22}, the challenge is to prepare samples that are dilute enough for SAXS while compact enough for XRD signal analysis.  

In monodisperse systems, the size information known as radius of gyration is often obtained from the SAXS curve slope near $Q=0$, above the half intensity and within the so-called Guinier region \cite{ag55}. In polydisperse systems, it has been emphasized here that the behaviour of the scattering peak FWHM as a function of NP size dispersion is summarized by ${\rm FWHM}^{-1} \propto \widetilde{L}_6$. However, going into a detailed discussion about the radius-of-gyration behaviour with the size dispersion is beyond the scope of this work as it can be different than the $\widetilde{L}_6$ behaviour \cite{sm22}. Shape dispersivity can also be addressed via X-ray scattering simulation in virtual NPs. For instance, NPs of rectangular shape are seen among the cubic ones in the SEM images, Fig.~\ref{fig:plotB5B11C1}(b-d). A virtual system of cubic and rectangular NPs can be used to check whether it explains diffraction peaks of systematically narrower-than-average widths, as observed for the \{200\} and \{400\} reflections, peaks number 2 and 6 in Fig.~\ref{fig:fwhmB5B11C1}(a). Beyond the basic illustrative demonstrations made in this work about the correlations between median values and peak widths, the role of polydispersivity on X-ray scattering can be further investigated via simulation in virtual NPs as a relatively simple procedure. In general, theoretical approaches that have been deduced mostly based on monodisperse systems, such as the asymptotic behavior of the SAXS curve as a function of $Q$ \cite{ag55,tz02,zl13}, can be tested in well controlled polydisperse systems instead of actual systems where dispersivity properties are difficult to control.

\section{Conclusions}

Comparing SAXS and XRD size results brings as benefit possible procedures for attesting the reliability of both bulk techniques in accessing size distribution parameters. Particularly in this work, the SAXS intensity curves of ceria powders were precisely reproduced by fitting functions based on polydisperse sizes of non-interacting NPs. Only after introducing XRD size results, there was new information to question the reliability of the SAXS curve fitting procedure. It revealed that the presence of NP-NP interaction effects in the SAXS curves can compromise assertive assessment of the NPs' properties. Consequently, being able to prepare dilute samples for SAXS measurements is a preliminary requirement for analyzing size dispersivity and crystallinity of NPs via a combined SAXS/XRD methodology. The XRD apparent crystallite size from Scherrer equation has to be compared to the median value of the size distribution fourth-moment integral from the SAXS curve fitting, a fact that was duly highlighted by the XRD simulation in polydisperse systems of virtual NPs. Likewise, SAXS simulation in virtual NPs has shown that measurements of the SAXS curve width in sufficiently dilute samples lead to the median value of the size distribution sixth-moment integral as the apparent size result. Broad size dispersivity is also a cause of discrepancies in  XRD and SAXS apparent sizes, implying that experimental observation of small discrepancies can be taken as an evidence of highly crystalline NPs with narrow size dispersivity. The relationship provided in this work to correlate discrepancies of apparent sizes was based on lognormal probability distribution functions as valid for describing both the particle sizes and crystalline domain sizes at the same time. It is a first step towards more general situations that can be exploited by combining SAXS and XRD techniques. 

\begin{acknowledgements}
A.V., R.F.S.P., M. B. E., and S.L.M. acknowledges financial support from CAPES (finance code 001), FAPESP (Grant No. 2019/15574-9, 2019/01946-1, 2021/01004-6), and CNPq (Grant No. 310432/2020-0). Thanks are due to University of S{\~{a}}o Paulo (NAP-NN) for funding the XRD equipment used in this research. We also acknowledge A. G. Oliveira-Filho from the Complex Fluids Group for technical assistance with SAXS measurements. F.J.T., B.C.R., S.D., and A.S.F acknowledges the Multiuser Central Facilities of UFABC, as well as support from the Center for Innovation on New Energies Shell (ANP)/FAPESP (2017/11937‐4) and Brazilian CNPq.
\end{acknowledgements}

\bibliography{manuscriptCGDout2022_arXiv}
\clearpage
\onecolumngrid
\appendix

\makeatletter
\renewcommand{\thefigure}{S\@arabic\c@figure}
\renewcommand{\thetable}{S\@arabic\c@table}
\renewcommand{\theequation}{S\arabic{equation}}
\renewcommand{\thepage}{S\arabic{page}}
\makeatother

\setcounter{figure}{0} 
\setcounter{page}{1}

\section{Supporting Information}\label{sec:appendix}

\section*{S1 - ATOMIC DISORDER}\label{sec:S1}

\begin{figure}
    \centering
    \includegraphics[width=\textwidth]{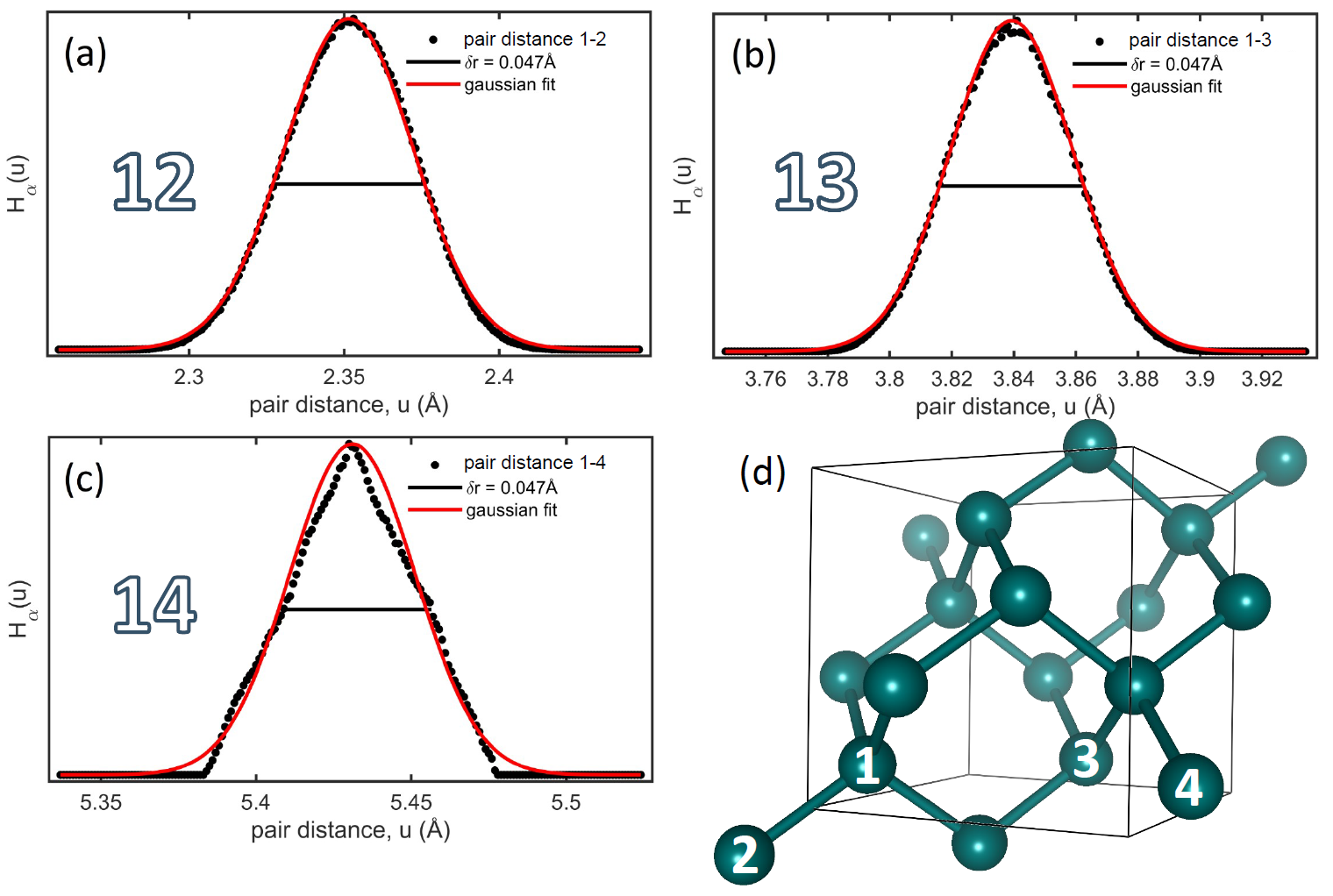}
    \caption{(a-c) Broadening in the histogram $H_\alpha(u)$ of pair distances caused by atomic vibration with RMS displacement $\langle dr \rangle_{\rm rms} = 0.3\delta r = 0.014\,\mathring{\rm A}$ in a silicon unit cell.  The histograms (scatter dots) were computed with a bin of $0.001\,\mathring{\rm A}$ within an ensemble of 1000 unit cells with random disorder of 2\%, that is $\delta r = 0.047\,\mathring{\rm A}$. Gaussian fits (solid red lines) with standard deviation $\sigma =  0.02\,\mathring{\rm A}$, Eq.~(\ref{eq:sigma}), are also shown. (d) Silicon unit cell. Atoms indicated by numbers were used to calculate the pair distances $r_{12}$, $r_{13}$, and $r_{14}$ along the [111], [110], and [100] directions, respectively. }
    \label{fig:pduwidth}
\end{figure}

Vibrational disorder has been treated within the frozen-in-time approach of the Debye scattering equation (DSE) \cite{Apd15,Aps16,Asm16}, as X-ray diffraction of randomly oriented identical NPs can be exactly calculated by using the DSE. For materials made of a single element $\alpha$, such as silicon NPs, the DSE is simply 
\begin{equation}\label{eq:intpofu}
    P(Q) = \lvert f_\alpha(Q)\rvert ^2\int H_{\alpha}(u)\frac{\sin(Qu)}{Qu} du
\end{equation}
where $Q=(4\pi/\lambda)\sin\theta$ is the reciprocal vector modulus for a scattering angle $2\theta$ and X-rays of wavelength $\lambda$. The pair distance distribution function (PDDF)
\begin{equation}\label{eq:hofu} 
    H_{\alpha} (u) = N_\alpha \delta(u) + 2\sum_{a = 1}^{N_\alpha} \sum_{b>a}^{N_\alpha} \delta (u - r_{ab})
\end{equation}
is the histogram of atomic distances $r_{ab} =  \lvert\vec{r}_b-\vec{r}_a\rvert$ for NPs with $N_\alpha$ atoms; $\delta()$ stands for the Dirac delta function. The atomic scattering factor  $f_\alpha(Q)=f_0(Q)+f^{\prime}(\lambda)+if^{\prime\prime}(\lambda)$ in Eq.~(\ref{eq:intpofu}) have been calculated by routines \texttt{asfQ.m} and \texttt{fpfpp.m}, both available at \citeauthor{Amc16}(\citeyear{Amc16})\cite{Amc16}. Small disorder in the atomic positions ${\vec r}_a = \langle {\vec r}_a \rangle + d\vec r$ were generated by adding $d\vec{r}= \delta r [ \zeta_1,\,\zeta_2,\,\zeta_3]$ to the mean atomic positions $\langle \vec{r}_a\rangle = [X_a,\,Y_a,\,Z_a]$ of the crystal lattice. The random numbers $\zeta_n$ are in the range $[0,1]$. The atomic disorder $\delta r$ produces a broadening of width $\delta r$ in the histograms of pair distances, as illustrated in Fig.~\ref{fig:pduwidth} by disordering a single unit cell several times. A Gaussian of standard deviation
\begin{equation}\label{eq:sigma}
   \sigma = \frac{\delta r}{2\sqrt{2\ln(2)\,}} = \sqrt{2\langle dr \rangle_{\rm rms}^2\,}    
\end{equation}
fits very well this broadening, corresponding to an isotropic root-mean-square (RMS) displacement $\langle dr \rangle_{\rm rms} = \delta r/(4\sqrt{\ln{2}\,}) \simeq 0.3\delta r$ around the mean atomic positions. 

\begin{figure}
    \centering
    \includegraphics[width=.9\textwidth]{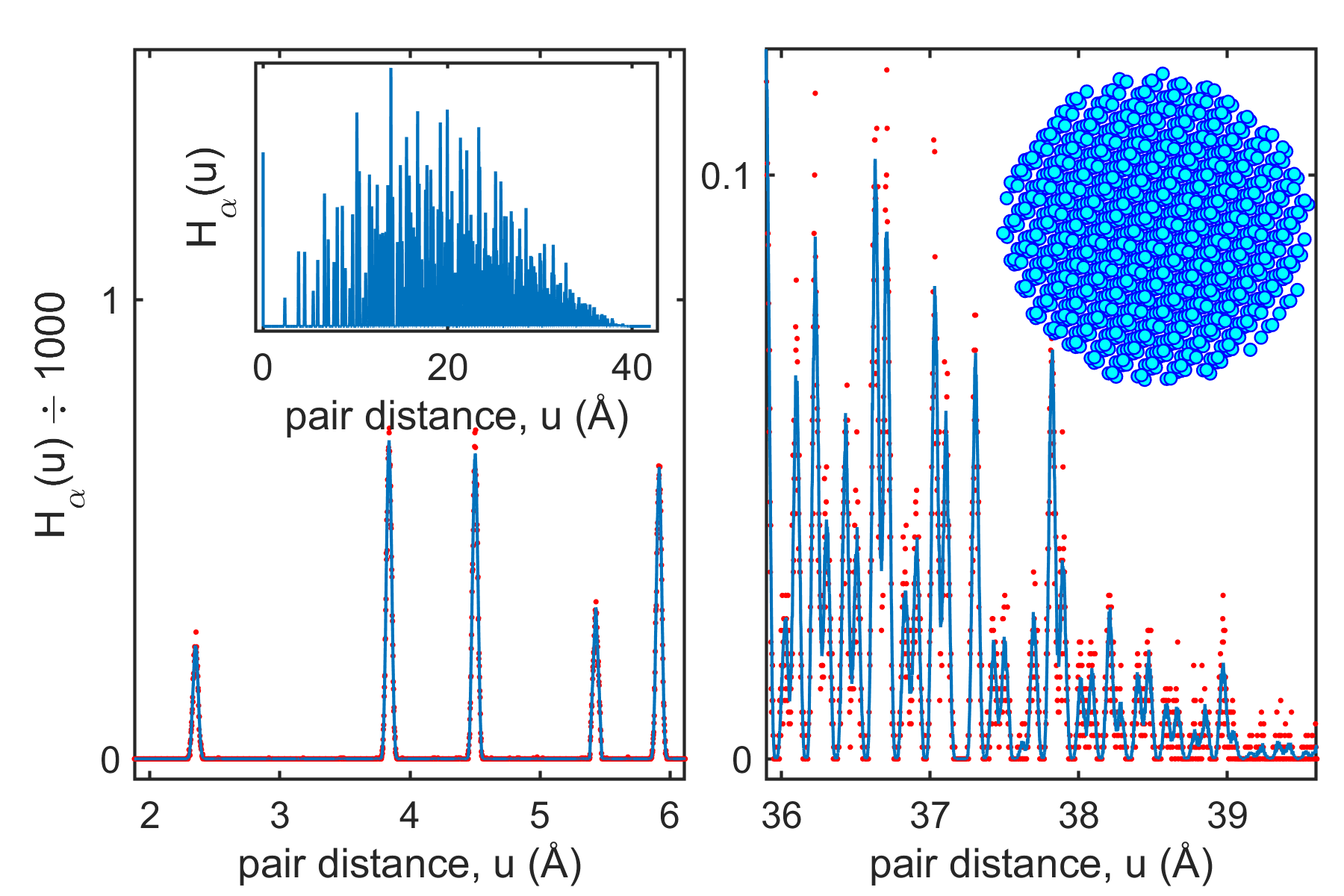}
    \caption{PDDF (red dots) of a 2\%-disordered 4\,nm diameter silicon NP  compared to the average PDDF (solid blue line). Left panel: pair distances of the first neighbors. Right panel: longest distances in the NP. PDDF of the whole NP, as well as the NP itself are shown as insets. All distances are seen with a standard deviation $\sigma = 2$\,pm. Total number of atoms in the NP: $H_\alpha(0) = 1672$ (inset, left panel). Bin width of $0.002\,\mathring{\rm A}$. }
    \label{fig:siliconD40}
\end{figure}

\begin{figure}
    \centering
    \includegraphics[width=\textwidth]{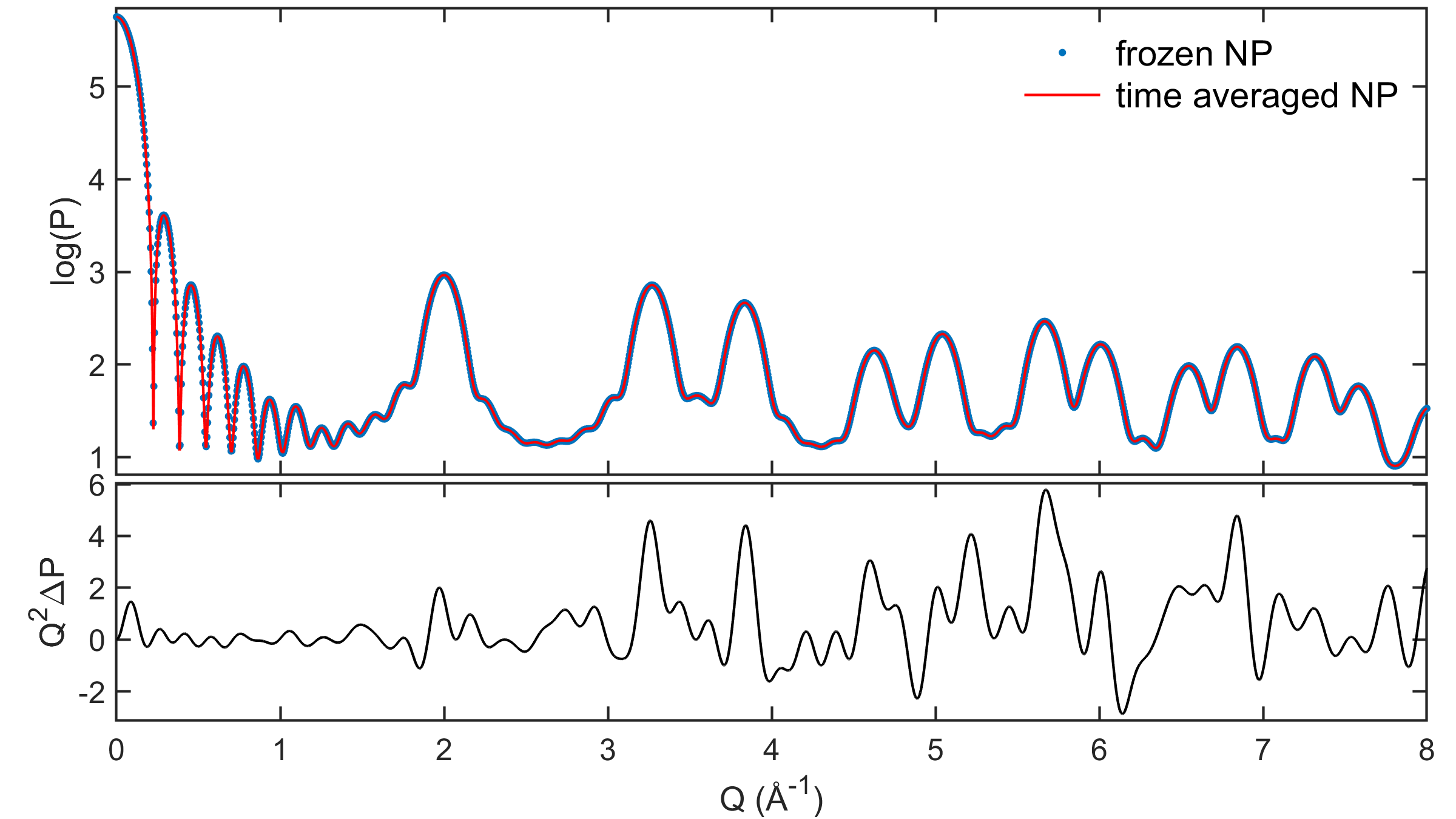}
    \caption{Top panel: Scattering power $P(Q)$, Eq.~(\ref{eq:intpofu}), of a single 2\%-disordered 4\,nm diameter silicon NP (blue dots) in comparison to the average scattering from an ensemble of disordered NPs (solid red line). Bottom panel: Difference $\Delta P$ of the above curves multiplied by $Q^2$ to make visible on linear scale tiny differences in the wide-angle region ($Q>1\,\mathring{\rm A}^{-1}$). }
    \label{fig:xrdpdu40}
\end{figure}

In NPs with sizes above a few nanometers, as the 4\,nm diameter NP in Fig.~\ref{fig:siliconD40}, the PDDF obtained for a single $\delta r$-disordered NP is very similar to the average PDDF from an ensemble of NPs. For the shortest distances the PDDFs are almost identical, Fig.~\ref{fig:siliconD40} (left panel), while for the longest distances the poor statistic from a single NP is more evident, as can be seen in Fig.~\ref{fig:siliconD40} (right panel). However, the NPs scattering power $P(Q)$, that is the XRD patterns from systems of randomly oriented NPs, exhibit tiny irrelevant differences only, as compared in Fig.~\ref{fig:xrdpdu40}. For larger NPs, bulk properties dominates  and the poor statistic of the longest distances are even more irrelevant. Therefore,  for the purpose of XRD simulation, instead of concerning with the average PDDF 
\begin{equation}\label{eq:Hmean}
   \langle H_{\alpha} (u)\rangle = H_{\alpha} (u)\ast G(u) = N_\alpha \delta(u) + 2\sum_{a = 1}^{N_\alpha} \sum_{b>a}^{N_\alpha} G(u - r_{ab})\,,
 \end{equation}
given by the convolution of Eq.~(\ref{eq:hofu}) with a Gaussian function $G(u)$ of unit area and standard deviation $\sigma$, Eq.~(\ref{eq:sigma}), it is enough to compute the PDDF of a single $\delta r$-disordered NP.  

\newpage

\section*{S2 - LINE PROFILE ANALYSIS}

\begin{figure}
    \centering
    \includegraphics[width=\textwidth]{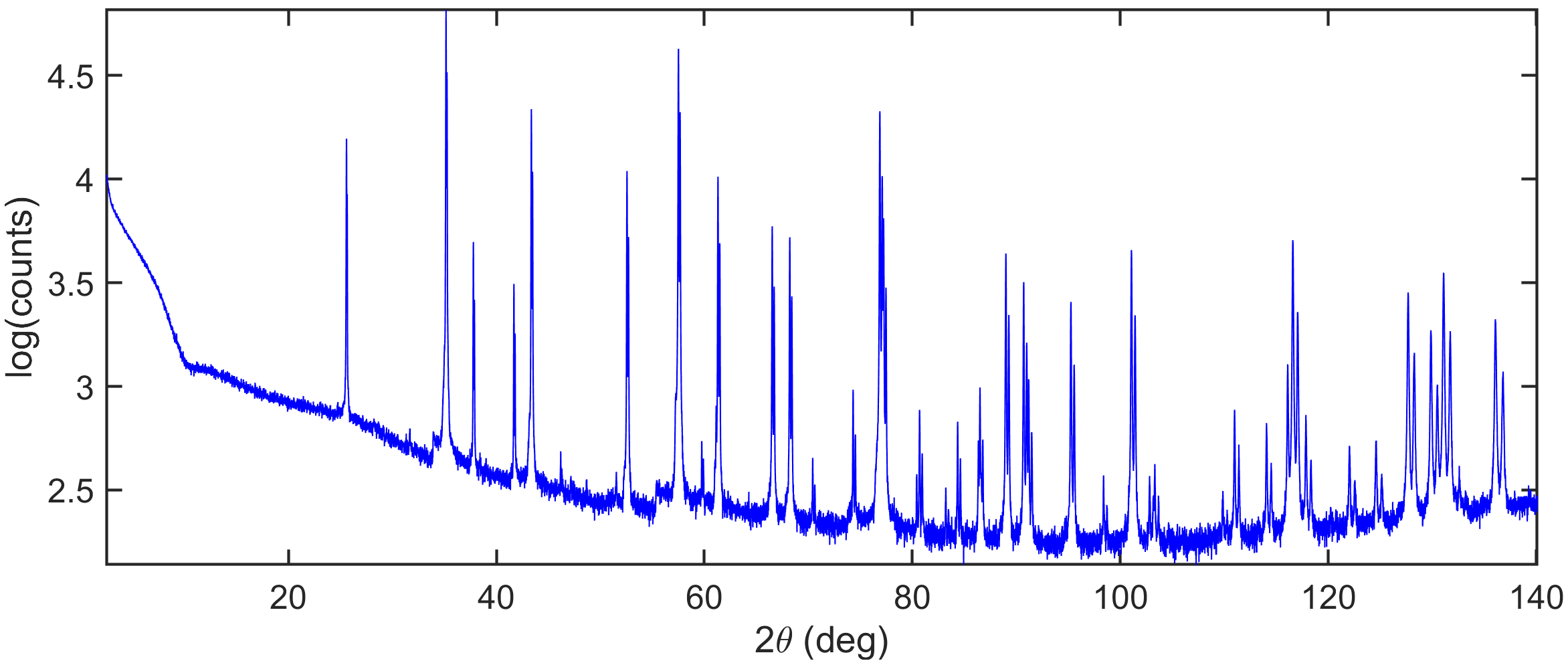}
    \caption{XRD pattern of a corundum NIST standard in a D8 Discover Bruker diffractometer. Cu$K_\alpha$ radiation (Ni filter). }
    \label{fig:xrdkorundum}
\end{figure}

Diffraction peak widths, that is FWHM values, were measured by line profile fitting of the doublet peaks due to  Cu$K_\alpha$ radiation. It was  carried out by a pseudo-Voigt double function $V(2\theta) = V_1(2\theta)+0.485V_2(2\theta)$ where 
\begin{equation}\label{eq:Yn}
    V_n(2\theta) = x G(2\theta - 2\theta_n) + (1-x) L(2\theta - 2\theta_n)\,.
\end{equation}
$G$ and $L$ stand for Gaussian and Lorentzian functions, respectively, both of height $h$. $\theta_n$ is the Bragg angle for an atomic interplane distance $d$ and walength $\lambda_n$ ($\lambda_1 =1.540562$\,\AA\, and $\lambda_2 =1.544426$\,\AA). For each hkl reflection family, the adjustable parameters are $x$, $h$, $d$, and the widths $w_G$ and $w_L$ of the $G$ and $L$ functions. Background fitting is based on linear interpolation of the baseline intensity near the peaks, as detailed elsewhere. \cite{Aac19} Data fittings were driven by a genetic algorithm that minimizes the mean-square error of the data \cite{Awor99}.    

For the \ce{CeO2} powder samples, the fitting of the diffraction peaks with the $V(2\theta)$ function provide the experimental width $\mathcal{W}_{\rm exp}$. Examples of fitting are given in the main text, Fig.~7(b-d). The general relationship 
\begin{equation}
    {\rm FWHM}(2\theta) = \frac{2\pi}{\lambda}\cos(\theta) {\rm FWHM}(Q)
\end{equation}
have been used to convert between the FWHM values of diffraction peaks observed in intensity curves plotted either as a function of $2\theta$ or $Q$.

\begin{figure}
    \centering
    \includegraphics[width=\textwidth]{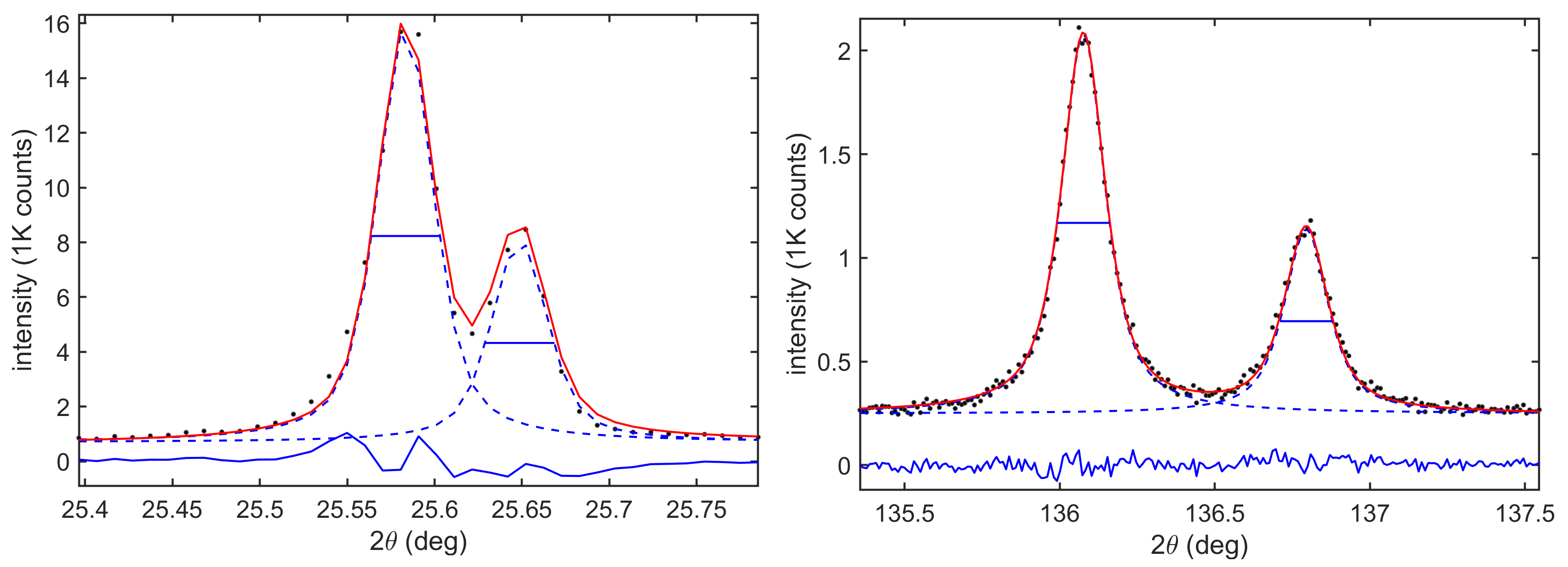}
    \caption{Examples of line profile fitting (solid red lines) of $K_{\alpha1}$ and $K_{\alpha2}$ diffraction peaks (scatter dots) by using the pseudo-Voigt double function $V(2\theta)$, Eq.~(\ref{eq:Yn}). The FWHM of each $V_n(2\theta)$ function (dashed lines) are indicated by horizontal solid lines. The residuals of curve fitting are also shown (blue solid lines). Both panels display Bragg reflections from the corundum XRD pattern in Fig.~\ref{fig:xrdkorundum}.}
    \label{fig:korundumpeaks}
\end{figure}

\begin{figure}
    \centering
    \includegraphics[width=.5\textwidth]{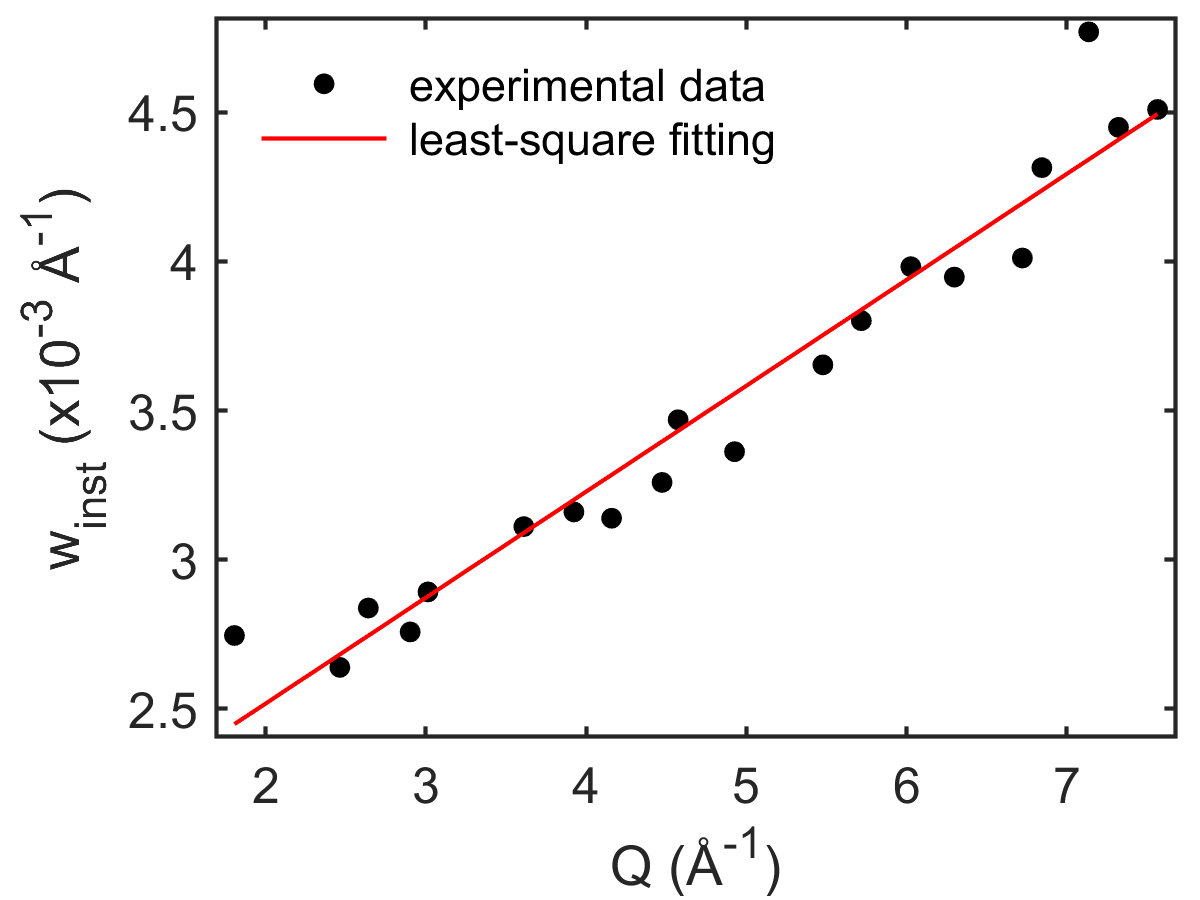}
    \caption{Instrumental width (scatter points) from corundum diffraction peak analysis. Least-square fitting (solid red line) of the experimental data leads to the instrumental width $\mathcal{W}_{\rm inst}(Q) = (0.355 Q + 1.8) \times 10^{-3}$\,\AA$^{-1}$ as a function of $Q=(4\pi/\lambda_1)\sin(\theta)$. }
    \label{fig:wsfunction}
\end{figure}

Instrumental brodening of diffraction peaks were determined from the XRD pattern of a standard corundum powder in Fig.~\ref{fig:xrdkorundum}. Examples of corundum diffraction peaks fitted by $V(2\theta)$ function are shown in Fig.~\ref{fig:korundumpeaks}. The FWHM values, $\mathcal{W}_{\rm inst}$, obtained from diffraction peak analysis are presented in Fig.~\ref{fig:wsfunction} as a function of the diffraction vector modulus $Q$. They were used to deconvolve the instrumental broadening that exist in $\mathcal{W}_{\rm exp}$, leading to the instrumental free peak width $\mathcal{W}=\left(\mathcal{W}_{\rm exp}^2-\mathcal{W}_{\rm inst}^2\right)^{1/2}$ of the \ce{CeO2} diffraction peaks.  

\begin{figure}
    \centering
    \includegraphics[width=.75\textwidth]{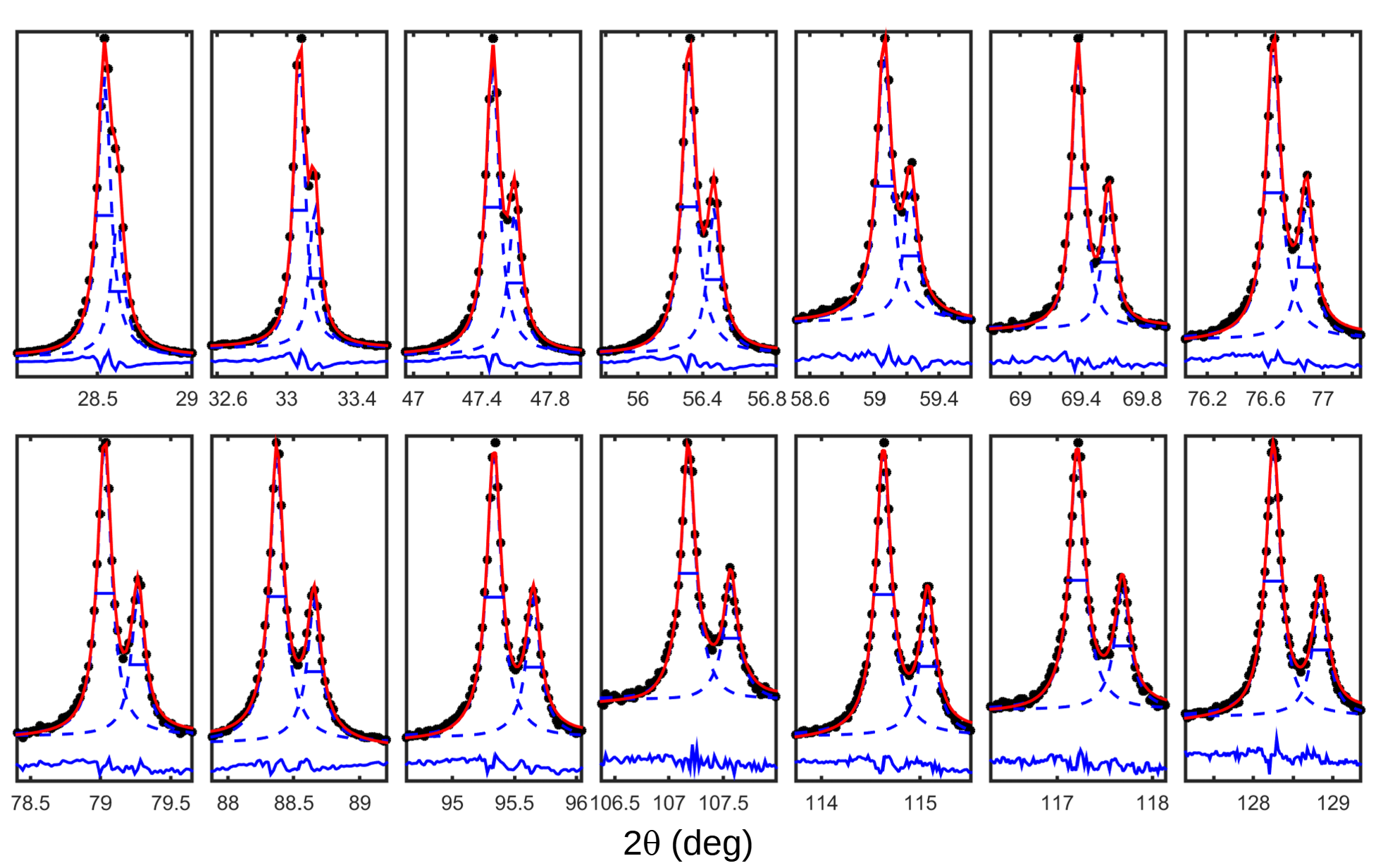}
    \caption{XRD of sample C1. Line profile fitting (solid red lines) of $K_{\alpha1}$ and $K_{\alpha2}$ diffraction peaks (scatter dots) by using the pseudo-Voigt double function $V(2\theta)$, Eq.~(\ref{eq:Yn}). The FWHM of each $V_n(2\theta)$ function (dashed lines) are indicated by horizontal solid lines. The residuals of curve fitting are also shown (blue solid lines).}
    \label{fig:C1}
\end{figure}

\begin{figure}
    \centering
    \includegraphics[width=.75\textwidth]{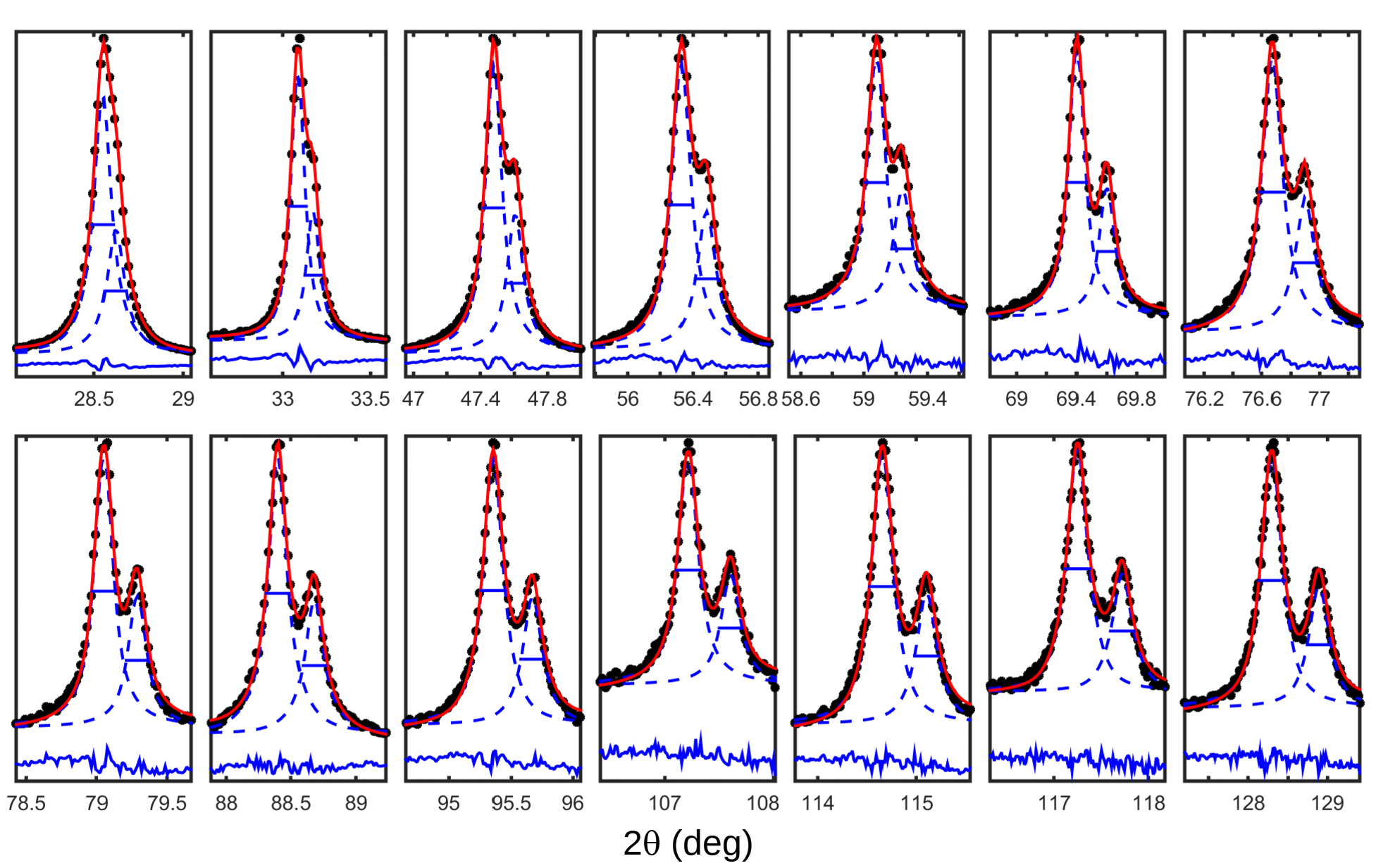}
    \caption{XRD of sample B5. Line profile fitting (solid red lines) of $K_{\alpha1}$ and $K_{\alpha2}$ diffraction peaks (scatter dots) by using the pseudo-Voigt double function $V(2\theta)$, Eq.~(\ref{eq:Yn}). The FWHM of each $V_n(2\theta)$ function (dashed lines) are indicated by horizontal solid lines. The residuals of curve fitting are also shown (blue solid lines).}
    \label{fig:B5}
\end{figure}

\begin{figure}
    \centering
    \includegraphics[width=.75\textwidth]{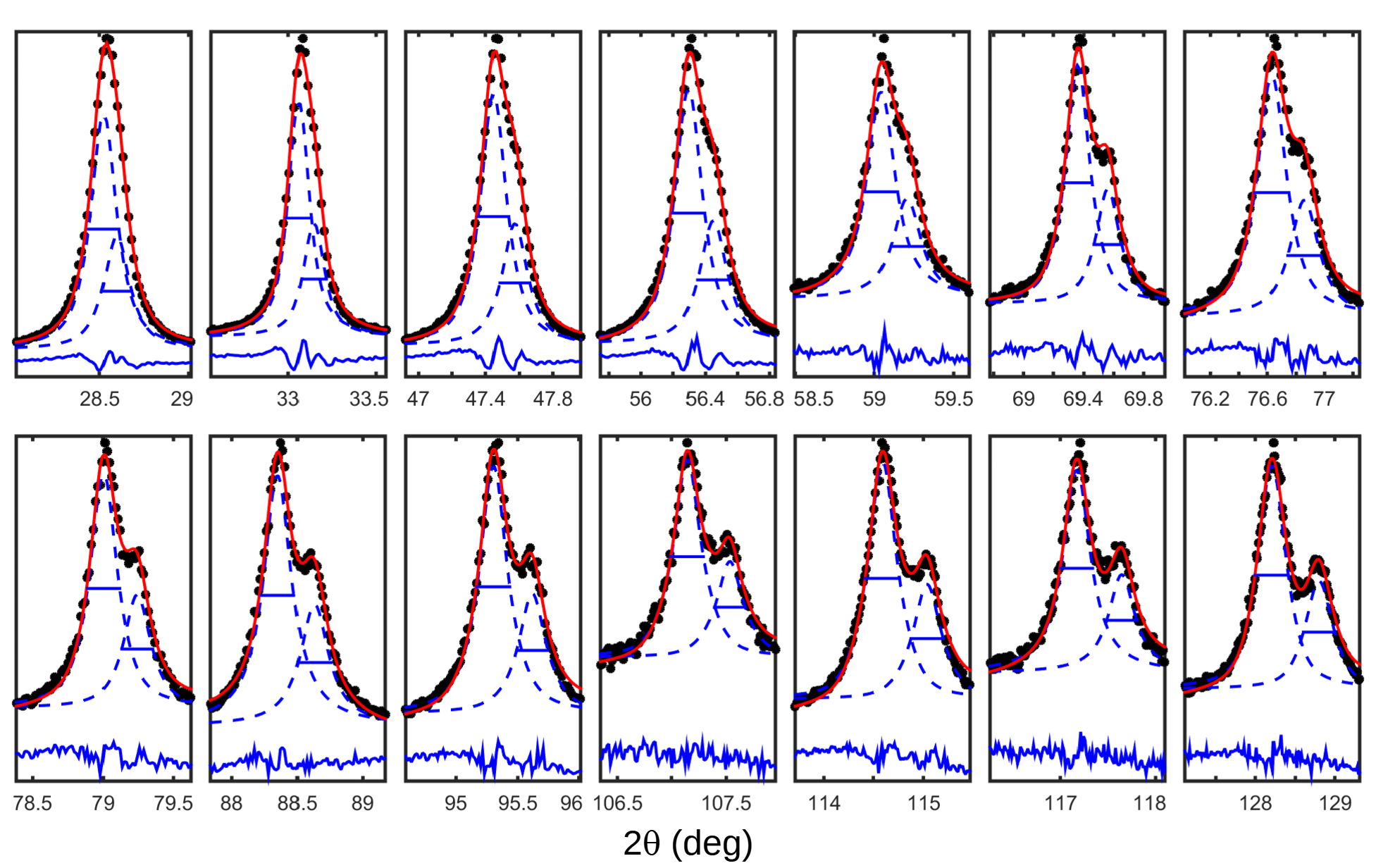}
    \caption{XRD of sample B11. Line profile fitting (solid red lines) of $K_{\alpha1}$ and $K_{\alpha2}$ diffraction peaks (scatter dots) by using the pseudo-Voigt double function $V(2\theta)$, Eq.~(\ref{eq:Yn}). The FWHM of each $V_n(2\theta)$ function (dashed lines) are indicated by horizontal solid lines. The residuals of curve fitting are also shown (blue solid lines).}
    \label{fig:B11}
\end{figure}

The experimental XRD peak widths $\mathcal{W}_{\rm exp}$ for samples C1, B5, and B11 were determined by line profile fitting with function Eq.~(\ref{eq:Yn}), as presented in Fig.~\ref{fig:C1}, Fig.~\ref{fig:B5}, and Fig.~\ref{fig:B11}, respectively. 


\clearpage
\section*{S3 - WEIGHTED LOGNORMAL FUNCTION}

In the particular case of a lognormal PSD function 
\begin{equation}\label{eq:ln}
    n(k)=\frac{1}{k \sigma \sqrt{2\pi}}\exp\left[-{\frac{\ln^2(k/\tilde{k})}{2\sigma^2}}\right]\,,
\end{equation}
the median value $\widetilde{K}_m$ of the $k^m$-weighted PSD is obtained as follow.

Integral solution by variable substitution is applicable for the integral 
\begin{equation}\label{eq:AK}
    A_1=\int_0^{\widetilde{K}_m} k^m n(k) dk = \frac{1}{\sigma \sqrt{2\pi}}\int_0^{\widetilde{K}_m}k^{(m-1)}\exp\left[-{\frac{\ln^2(k/\tilde{k})}{2\sigma^2}}\right]dk
\end{equation}
by changing $u = \sigma^{-1} \ln(k/\tilde{k})$, $du = (k\sigma)^{-1} dk$, and the upper and lower integral limits by $\tilde{u} = \sigma^{-1}  \ln(\widetilde{K}_m/\tilde{k})$  and $-\infty$, respectively. Then, 
\begin{equation}\label{eq:Au}
    A_1=\frac{\tilde{k}^m}{\sqrt{2\pi}}\int_{-\infty}^{\tilde{u}}{\rm e}^{(m u \sigma - u^2/2)}du = \frac{\tilde{k}^m {\rm e}^{m^2\sigma^2} }{\sqrt{2\pi}}\int_{-\infty}^{\tilde{u}} 
    {\rm e}^{-\frac{1}{2}(u-m\sigma)^2}du\,.
\end{equation}
Substitution of variable can be applied one more time, as $a = (u-m\sigma)/\sqrt{2}$ and $da = du/\sqrt{2}$, for which the upper and lower limits become $\tilde{a} = (\tilde{u}-m\sigma)/\sqrt{2}$ and $-\infty$. It results in
\begin{equation}\label{eq:AKleft}
    A_1 = \frac{\tilde{k}^m {\rm e}^{m^2\sigma^2} }{\sqrt{\pi}}\int_{-\infty}^{\tilde{a}} 
    {\rm e}^{-a^2}da = 
    \frac{1}{2}\tilde{k}^m {\rm e}^{m^2\sigma^2} \left[{\rm erf}(\tilde{a})+1\right] 
\end{equation}
where erf is the the error function \cite{Ajg71,Ala97}. As $\widetilde{K}_m$ stands for the median value, the integral 
\begin{equation}\label{eq:AKright}
    A_2=\int_{\widetilde{K}_m}^\infty k^m n(k) dk = \frac{1}{2}\tilde{k}^m {\rm e}^{m^2\sigma^2} \left[-{\rm erf}(\tilde{a})+1\right]
\end{equation}
can be solve by the same procedure of $A_1$, and both integrals must be equal. $A_1=A_2$ in Eqs.~(\ref{eq:AKleft}) and (\ref{eq:AKright}) is possible as long as $\tilde{a}=0$, or $\tilde{u}=\sigma^{-1}\ln(\widetilde{K}_m/\tilde{k})= m\sigma$. It implies that 
\begin{equation}\label{eq:K}
   \widetilde{K}_m = \tilde{k}\exp(m \sigma^2)= k_0\exp[(m+1) \sigma^2]\,,
\end{equation}
as used in this work for the $k^4$- and $k^6$-weighted PSD. $\tilde{k} = k_0\exp(\sigma^2)$ is the unweighted PSD median value written in terms of the PSD mode $k_0$ and standard deviation in log scale $\sigma$.

\providecommand{\latin}[1]{#1}
\makeatletter
\providecommand{\doi}
  {\begingroup\let\do\@makeother\dospecials
  \catcode`\{=1 \catcode`\}=2 \doi@aux}
\providecommand{\doi@aux}[1]{\endgroup\texttt{#1}}
\makeatother
\providecommand*\mcitethebibliography{\thebibliography}
\csname @ifundefined\endcsname{endmcitethebibliography}
  {\let\endmcitethebibliography\endthebibliography}{}

\end{document}